\documentclass[fleqn,usenatbib,onecolumn]{mnras}

\usepackage{newtxtext,newtxmath}
\usepackage[utf8]{inputenc}
\usepackage[T1]{fontenc}

\DeclareRobustCommand{\VAN}[3]{#2}
\let\VANthebibliography\thebibliography
\def\thebibliography{\DeclareRobustCommand{\VAN}[3]{##3}\VANthebibliography}

\usepackage{graphicx}	% Including figure files
\usepackage{amsmath}	% Advanced maths commands
\usepackage{esint}
\usepackage[OT1]{fontenc}

\allowdisplaybreaks[1]
\title[Induction Heating]{Stellar-Field-Driven Induction Heating Unlikely To Cause Significant Heating in Giant Exoplanets}

\author[M. Brady et al.]{
Madison Brady,$^{1}$\thanks{E-mail: bradym27@msu.edu}
Cassidy  Walker,$^{1}$
Darryl Z. Seligman,$^{1}$
\\
$^{1}$ Dept. of Physics and Astronomy, Michigan State University, East Lansing, MI 48824, USA \\
}

\date{Accepted XXX. Received YYY; in original form ZZZ}

\pubyear{2026}

\begin{document}
\label{firstpage}
\pagerange{\pageref{firstpage}--\pageref{lastpage}}
\maketitle

\begin{abstract}
A planet moving through a changing magnetic field will have a current induced on it by that magnetic field. If the magnetic Reynolds number in the upper atmosphere of this planet is small, the dissipation of this current could result in observable changes to the planet's properties. In this paper, we estimated the total power that would be dissipated in the observed giant exoplanet population if they orbited stars with misaligned magnetic dipoles.  We found that, even if all planets orbit stars with extremely misaligned surface magnetic fields, the vast majority of known exoplanets are unlikely to dissipate enough magnetic power to cause observable changes in their surface temperatures.  We also extended our calculations to estimate the impact that a low-but-nonzero planetary eccentricity would have on the induced power, and provide a code that implements these calculations. Overall, while it is unlikely that any currently known exoplanets are experiencing a high enough degree of stellar-induced Ohmic heating to be observable, we cannot rule out the possibility that an eccentric, inclined planet on a short-period orbit around a superlatively magnetically active star could have observable magnetic heating. We note that our calculations assume that magnetic diffusion dominates over induction in the upper atmospheres of these planets.  However, hot Jupiters with large wind speeds may have high magnetic Reynolds numbers and therefore be in the inductive regime.
\end{abstract}


\section{Introduction}

%\textcolor{red}{bullet points just list some general notes I want to hit}

An object moving through a time-varying magnetic field will have a current induced on its surface \citep{Jackson1962}. Following this logic, it is reasonable to assume that a planet orbiting a star with a misaligned magnetic field may experience an induced current, which could give rise to planetary heating through Ohmic dissipation. Planets on eccentric orbits may experience similar effects.

Planets in the Solar System are largely aligned \citep{Laskar1988,Casanovas1997,Beck2005}; however, observations of exoplanetary systems have displayed a broader range of orbital properties \citep{Winn2015}. There are many giant planets with heightened eccentricities or misaligned orbits \citep{Albrecht2022, Albrecht2012}. \citet{Gilbert2025} showed that planets with radii larger than about 3.5\,$R_\oplus$ have an average population eccentricity of 0.2, while \citet{Morgan2026} showed that the majority of warm Jupiters appear to be dynamically hot. In addition, Jupiters on orbits misaligned with their host stars' rotation axes appear to also be fairly common, especially when considering planets orbiting hot stars \citep{Winn2010}. \citet{Wang2026}, meanwhile, showed that Jupiter-mass exoplanets can be misaligned or eccentric, but typically not both.

Multiple theories have been proposed to explain the prevalence of planets on eccentric or misaligned orbits. In multi-planet systems, planet-planet scattering \citep{Chatterjee2007, Ford2008, Juric2008} and mean-motion resonances \citep{Lee2002, Thommes2003} can alter planetary orbits, either by increasing eccentricity or forcing a mutual inclination. Additionally, planets can be perturbed onto more misaligned or eccentric orbits by distant external companions via the Kozai-Lidov mechanism \citep{Kozai1962, Lidov1962, Naoz2016}. These various gravitational interactions have also been used to explain the observed populations of close-in giant planets and hot Jupiters \citep{Dawson2018, Fabrycky2007}.

Some hot Jupiters exhibit larger radii than can be explained with stellar irradiation alone \citep{Baraffe2010,Deming2009,Miller2009}. Ohmic dissipation in some form has frequently been invoked to explain this radius inflation. For example, Ohmic dissipation caused by the interaction between zonal winds and planetary magnetic field has been shown to cause significant heating in hot Jupiters \citep{Batygin2010, Batygin2011, Perna2010, Perna2010b, Menou2012b}. However, this mechanism only operates efficiently to inflate hot Jupiter atmospheres at the initial phase before the planet cools \citep{Wu2013}.%, and therefore other effects have been investigated to explain the distributions of older Hot Jupiter radii \citep{Laughlin2011,Owen2014}. 

Moreover, \citet{Menou2012} argued that a thermoresistive instability could operate in these hot Jupiter atmospheres, where if the Ohmic heating was more efficient than re-radiation, a positive feedback loop between heat deposited via induced currents and the temperature dependence of conductivity would lead to a runaway. This mechanism was also investigated in further detail with 3-dimensional global circulation considerations \citep{Rauscher2013,Beltz2022} and with time-evolving conductivity \citep{Hardy2022}. However, these models all consider the Ohmic heating induced by the planetary magnetic field, as opposed to the stellar magnetic field.

Ohmic dissipation has been investigated as a driver of planetary heating via interaction between the planet and stellar magnetic fields. For example, \citet{Laine2008} have demonstrated that the power deposited onto planets via this method has the potential to cause radius inflation, Roche lobe overflow, and even influence orbital migration for giant planets. These mechanisms were expanded in further detail by \citep{Chang2012} to include the effect of stellar obliquity and various interior structures. The resulting mass loss and orbital evolution was invoked to halt Type II \citep{Lin1996,Ward1997} migration scenarios. %However, their work only considers planets with circular orbits, and does not compare their model to observed results on published exoplanets.

%This mechanism has also been investigated as a source of heating and atmospheric escape for smaller, rocky planets. \citet{Cohen2024} investigated the effect of Joule heating from short-period rocky planets, and estimated that this effect could drive significant atmospheric escape for Trappist-1e and planets similar to this. A similar investigation was also applied to Trappist-1b and Pi Men c, demonstrating that Ohmic heating could be extremely efficient in these types of planets but depended critically on the conductivity profile of the atmosphere \citep{Strugarek2025}. Ohmic dissipation via interaction withe the stellar wind \citep{Cohen2014,Cohen2018} and  atmospheric escape  effects could shape the overall habitability of rocky planets \citep{Chatterjee2026}. It also has been demonstrated that for smaller rocky planets similar effects can produce observable effects on the star and excess heating of the planet, and potentially circularization and synchronization of the orbit \citep{Laine2012}. %These mechanisms were invoked to overcome or halt Type I \citep{Tanaka2002} and 

The launch of \textit{JWST} has given us a unique opportunity to probe the emission spectra of many large, close-in planets \citep[see, e.g.,][]{Morel2025, Saha2026}. This is a fantastic opportunity to measure, in-depth, the actual temperatures of these bodies, giving us the opportunity to study the energy balance on these planets, which can be influenced by properties such as the planet's albedo, atmospheric heat redistribution, and the internal heat from formation \citep[see, e.g.,][]{Mehta2026}. If heating due to magnetism-induced currents on these planets is significant, it may also result in a corresponding change in the planet's observed surface temperature, both 1) giving us an opportunity to probe the strength of this mechanism and 2) complicating our efforts at properly characterizing these bodies. It is therefore necessary to estimate the magnitude of induction heating on the known population of exoplanets, taking realistic values of stellar magnetic field magnitudes into consideration, to determine whether or not it is a significant contributor to the energy balance of these planets.

In Section~\ref{sec:methods}, we describe a model to estimate the amount of power deposited by induction heating on a short-period giant exoplanet in a low-Reynolds number regime, informed by the work from \cite{Laine2008} and \cite{Campbell1983}. In Section~\ref{sec:pind_mag}, we estimate the magnitude of the induced power $P_{ind}$ and show how system parameters such as the stellar magnetic dipole moment and the planetary orbital period influence the induced power. We also estimate the influence a nonzero eccentricity would have on our model and describe how to expand our calculations to higher-order in eccentricity, and briefly describe how planetary obliquity could be accounted for in our methods. In Section~\ref{sec:comp_to_exo}, we compare our calculated $P_{ind}$ values to the known exoplanet population, including bodies orbiting stellar remnants. In Section~\ref{sec:conc}, we conclude.

\section{Methods}
\label{sec:methods}

\subsection{Heating A Planet With A Changing Magnetic Field}
\label{ssec:heating_base}

We follow the methods from \cite{Laine2008} \citep[which, in turn, was inspired by][]{Campbell1983} to describe how power from a changing magnetic field is deposited inside of a planet. Consider a star with a magnetic dipole moment $m$ whose axis is tilted at an angle $\alpha$ relative to its rotation axis with a planet at an orbital separation $D$.  In this case, we assume the system has a zero obliquity - we refer the reader to \cite{Chang2012} for a discussion on how these calculations change in the event of a nonzero obliquity.  We also assume that the planet is synchronously rotating, such that the planet's rotation period is equal to its orbital period.   This is a reasonable assumption for planets on short-period orbits undergoing strong tidal synchronization \citep{Kasting1993}.

We are interested in determining the value of the magnetic field at an arbitrary point in space.  We define the coordinate system the same as in \cite{Campbell1983}, where the primed frame is centered on the primary star and the unprimed frame is centered on the planet.  We also note that $\varphi$ is the azimuthal angle and $\theta$ is the polar angle in this frame.

\subsubsection{The Magnetic Field Outside the Planet}
\label{sssec:mag_field}

We can start by calculating the scalar magnetic potential of the star in the frame of the star, $\Psi$:

\begin{equation}
    \Psi = \frac{m}{4\pi r'^3} \mathbf{r'} \cdot \mathbf{\hat{m}},
\end{equation}

\noindent where $\mathbf{\hat{m}}$ is the unit dipole moment, which is given by:

\begin{equation}
\label{eqn:mhat_defn}
    \mathbf{\hat{m}} = \sin(\alpha) \cos (\omega t) \mathbf{i'} + \sin(\alpha) \sin (\omega t) \mathbf{j'} + \cos(\alpha)  \mathbf{k'},
\end{equation}

\noindent and $\mathbf{r'}$ is the location of an arbitrary point in space near the planet, defined as:

\begin{equation}
    \mathbf{r'} = x' \mathbf{i'} +  y' \mathbf{j'} +  z' \mathbf{k'}.
\end{equation}

These calculations assume that the stellar magnetic field is a simple magnetic dipole.  We note that, outside the stellar Alfv\'{e}n radius \citep[about 10-20\,$R_\star$ for most stars; see][]{Boehm2025}, the stellar wind may drag and distort the shape of the magnetic field, and, within the source surface radius \citep[typically around 2-10 $R_\star$, see][]{Reville2015}, the stellar magnetic field loops may not be open.  The exact shape of the stellar magnetic field, Alfv\'{e}n surface, and source surface radius are complex and vary both in time and space \citep{Finley2025, Reville2015}, making it difficult to conclude anything about the exact form of the magnetic field.  The analysis in this section may therefore be an oversimplification with regards to planets that are extremely close to their host star or planets that orbit stars with complex multipolar magnetic fields \citep[such as accreting T Tauri stars, see][]{Gregory2008}.  However, we consider the dipolar field to be a reasonable assumption in most scenarios because (i) the dipolar component of the magnetic field decays the most slowly with distance in a vacuum and (ii) we have limited information regarding the stellar and planetary magnetic field topologies in most extrasolar systems.

We can therefore calculate $\Psi$ as:

\begin{equation}
\label{eqn:psi_defn}
    \Psi = \frac{m}{4\pi r'^3} \bigg(\sin(\alpha) \cos (\omega t)x' +  \sin(\alpha) \sin (\omega t) y' + \cos(\alpha) z'\bigg),
\end{equation}

\noindent where $\omega$ is the angular frequency of the rotating field; $\omega = 2\pi/p_{rot} - 2\pi/p_{orb}$, $p_{rot}$ is the stellar rotation period, and $p_{orb}$ is the planet's orbital period.

The potential in the frame of the planet can be calculated by performing these transformations:

\begin{equation}
\label{eqn:prime_values}
\begin{aligned}
    r' = \sqrt{r^2 + D^2 - 2rD \sin(\theta) \cos(\varphi)},~~~
    x' = r \sin(\theta) \sin(\varphi),~~~
    y' = D-r \sin(\theta) \cos(\varphi),~~~
    z' = r \cos(\theta).
\end{aligned}
\end{equation}

\noindent where $r$ is the distance between the center of the planet and the arbitrary point $(x', y', z')$.  If we are thinking about the magnetic field values at or near the surface of the planet, $r\lesssim R_p$.

Before we insert these values into Equation~\ref{eqn:psi_defn}, we will define a variable $\delta=r/D$.  As we anticipate $r/D$ to typically be small, this will enable us to perform some simplifications.  Firstly, we re-write these our transformations in terms of $\delta$:

\begin{equation}
    r' = D\sqrt{\delta^2 + 1 - 2\delta \sin(\theta) \cos(\varphi)},~~~
    x' = D\delta \sin(\theta) \sin(\varphi),~~~
    y' = D\bigg(1-\delta \sin(\theta) \cos(\varphi)\bigg),~~~
    z' = D\delta \cos(\theta).
\end{equation}

We can now write out $\Psi$ in the frame of the planet:

\begin{equation}
\begin{aligned}
    \Psi = \frac{m}{4\pi D^2} \times \bigg(\sin(\alpha) \cos (\omega t) \sin(\theta) \sin(\varphi) \frac{\delta}{\sqrt{\delta^2 + 1 - 2\delta \sin(\theta) \cos(\varphi)}^3} \\+  \sin(\alpha) \sin (\omega t)   \frac{\big(1-\delta \sin(\theta) \cos(\varphi)\big)}{\sqrt{\delta^2 + 1 - 2\delta \sin(\theta) \cos(\varphi)}^3} \\+ \cos(\alpha) \cos(\theta) \frac{\delta}{\sqrt{\delta^2 + 1 - 2\delta \sin(\theta) \cos(\varphi)}^3}\bigg).
\end{aligned}
\end{equation}

Assuming $\delta$ is small, we calculate out to second order in $\delta$:

\begin{equation}
\begin{aligned}
    \frac{\delta}{\sqrt{\delta^2 + 1 - 2\delta \sin(\theta) \cos(\varphi)}^3} \approx \delta + 3 \sin(\theta) \cos (\varphi) \delta^2,
\end{aligned}
\end{equation}

\noindent and:

\begin{equation}
\begin{aligned}
    \frac{\big(1-\delta \sin(\theta) \cos(\varphi)\big)}{\sqrt{\delta^2 + 1 - 2\delta \sin(\theta) \cos(\varphi)}^3} \approx 1 + 2 \sin(\theta) \cos(\varphi) \delta + \frac{3}{2}\big(3 \sin^2(\theta) \cos^2 (\varphi) - 1\big) \delta^2.
\end{aligned}
\end{equation}

Therefore, we can approximate $\Psi$ as:

\begin{equation}
\begin{aligned}
    \Psi \approx \frac{m}{4\pi D^2} \times \Bigg(\sin(\alpha) \cos (\omega t) \sin(\theta) \sin(\varphi) \big(\delta + 3 \sin(\theta) \cos (\varphi) \delta^2\big)\\+  \sin(\alpha) \sin (\omega t)   \bigg(1 + 2 \sin(\theta) \cos(\varphi) \delta + \frac{3}{2}\big(3 \sin^2(\theta) \cos^2 (\varphi) - 1\big) \delta^2 \bigg)\\+ \cos(\alpha) \cos(\theta) \big(\delta + 3 \sin(\theta) \cos (\varphi) \delta^2\big) \Bigg).
\end{aligned}
\end{equation}

Now, we can group factors of $\delta$ and find:

\begin{equation}
\begin{aligned}
    \Psi = \frac{m \sin(\alpha) }{4\pi D^2} \sin(\omega t)P_0^0 
    + \frac{m \sin(\alpha)r}{4\pi D^3} P_1^1 \bigg( \cos (\omega t) \sin(\varphi)  +  2 \sin (\omega t) \cos(\varphi) \bigg)
    + \frac{m \cos(\alpha) r}{4\pi D^3} P_1^0 \\
    -\frac{3 m \sin (\alpha) r^2}{8 \pi D^4} \Bigg(P_0^2\sin(\omega t)- P_2^2 \bigg( \frac{1}{2}\sin (\omega t)\cos (2\varphi) + \frac{1}{3}\cos (\omega t)  \sin(2\varphi)   \bigg) \Bigg)
    + \frac{m \cos(\alpha) r^2}{4\pi D^4} P_2^1 \cos (\varphi),
\end{aligned}
\end{equation}

\noindent where the $P_l^m$ are the Legendre polynomials:

\begin{equation}
\begin{aligned}
    P_0^0=1;~~
    P_1^0 = \cos(\theta);~~
    P_1^1 = \sin(\theta);~~
    P_2^0 = \frac{1}{2} \big(3\cos^2(\theta)-1);~~
    P_2^1 = 3 \sin(\theta) \cos(\theta);~~
    P_2^2 = 3\sin^2 (\theta).
\end{aligned}
\end{equation}

Assuming an eccentricity of zero (that is $D$ is constant), the time-dependent portion of this equation is:

\begin{equation}
\label{eqn:psi_grouped}
\begin{aligned}
    \Psi = \frac{ \mu_0 m \sin(\alpha) }{4\pi D^2} \sin(\omega t)P_0^0 
    + \frac{\mu_0 m \sin(\alpha)r}{4\pi D^3} P_1^1 \bigg( \cos (\omega t) \sin(\varphi)  +  2 \sin (\omega t) \cos(\varphi) \bigg)
    \\-\frac{3 \mu_0 m \sin (\alpha) r^2}{8 \pi D^4} \Bigg(P_0^2\sin(\omega t)- P_2^2 \bigg( \frac{1}{2}\sin (\omega t)\cos (2\varphi) + \frac{1}{3}\cos (\omega t)  \sin(2\varphi)   \bigg) \Bigg),
\end{aligned}
\end{equation}

\noindent which is identical to equation 6 in \cite{Campbell1983} besides a factor of $\mu_0$, which we include for comparison to \cite{Laine2008} and \cite{Chang2012}.

We can then find the poloidal scalar $\phi$ with the following relation:

\begin{equation}
\Psi = -\frac{\partial \phi}{\partial r}.
\end{equation}

Integrating Equation~\ref{eqn:psi_grouped}, we can find the magnetic field of the star at an arbitrary point in space near the surface of the planet:

\begin{equation}
\label{eqn:phi_star}
\begin{aligned}
    \phi_\star = -\frac{\mu_0 m \sin(\alpha) r}{4\pi D^2} \sin(\omega t)P_0^0 
    - \frac{\mu_0m \sin(\alpha)r^2}{8\pi D^3} P_1^1 \bigg( \cos (\omega t) \sin(\varphi)  +  2 \sin (\omega t) \cos(\varphi) \bigg)
     \\
    +\frac{ \mu_0m \sin (\alpha) r^3}{8 \pi D^4} \Bigg(P_0^2\sin(\omega t)- P_2^2 \bigg( \frac{1}{2}\sin (\omega t)\cos (2\varphi) + \frac{1}{3}\cos (\omega t)  \sin(2\varphi)   \bigg) \Bigg).
\end{aligned}
\end{equation}

Here, we can see that $\phi_\star$ is proportional to $m$, the magnetic dipole moment of the star, as well as sin($\alpha$), where $\alpha$ is the tilt of the star's spin axis relative to magnetic dipole.  If the dipole is aligned with the star's spin axis ($\alpha=0$), the time-dependent portion of $\phi_\star$=0.  $\phi_\star$ is also inversely proportional to $d$, the distance between the planet and the host star, meaning that, as expected, the planet ``feels'' significantly less magnetic field when it is very far from the host star.  

For future reference, the first term in $\phi_\star$ is a function of $P_0^0$, which has no spatial dependence.  As we will show later, the amount of power generation is based on the derivative of $\phi$ with respect to $\varphi$ and $\theta$.  Therefore, we can omit the $P_0^0$ term from our later calculations.

As discussed in \cite{Campbell1983}, the solutions of the form $r^{l+1}$ correspond to the poloidal scalars of the field from the star, while solutions of the form $r^{-l}$ correspond to poloidal scalars of the field due to the current inside of the planet.  Therefore, there is an accompanying set of solutions for $\phi_p$, the poloidal scalar of the field due to the induced current on the planet, that must also have the same time and angular dependence as $\phi_\star$, as it is induced by $\phi_\star$. Therefore, $\phi_p$ takes the form:

\begin{equation}
\label{eqn:phi_p}
\begin{aligned}
    \phi_p =  \frac{\mu_0}{r}  P_1^1\bigg(\alpha_1\sin(\omega t) + \alpha_2 \cos(\omega t)\bigg) \cos (\varphi) 
    + \frac{\mu_0}{r}  P_1^1\bigg(\alpha_3\sin(\omega t) + \alpha_4 \cos(\omega t)\bigg) \sin (\varphi)
    +\frac{\mu_0}{r^2} P_2^0 \bigg(\beta_1\sin(\omega t) + \beta_2 \cos(\omega t) \bigg) 
    \\+\frac{\mu_0}{r^2} P_2^2 \bigg( \gamma_1\sin(\omega t) + \gamma_2 \cos(\omega t)\bigg) \cos (2\varphi) 
    +\frac{\mu_0}{r^2} P_2^2 \bigg( \gamma_3\sin(\omega t) + \gamma_4 \cos(\omega t)\bigg) \sin(2\varphi) .
\end{aligned}
\end{equation}

With these, we now have a equation for the value of the poloidal scalar outside the planet, $\phi_{out} = \phi_\star+\phi_p$.  However, we next must consider what is occurring inside the planet.  

\subsubsection{The Magnetic Field Inside the Planet}

\cite{Laine2008} estimated the poloidal scalar $\phi_{inner}$ (assuming the field penetrates the planet quickly) with an expansion in spherical harmonics:

\begin{equation}
\label{eqn:phi_def}
    \phi_{inner} (r, t) = \mu_0 \bigg(\sum_{l, m} C^{m}_{l} G_l(r) Y_l^m(\theta, \varphi)\bigg)e^{i\omega t},
\end{equation}

\noindent where $C_l^m$ are constant coefficients, $G_l(r)$ is an arbitrary function $G\in\mathbb{C}$ that describes how the scalar evolves with $r$, and $Y_l^m(\theta, \varphi)$ are spherical harmonics.

\cite{Laine2008} calculated $G_l(r)$ inside the planet by considering that, from Maxwell's equations,

\begin{equation}
\label{eqn:G_def}
    \frac{d^2 G_l}{d r^2}(r) - \bigg(\frac{l(l+1)}{r^2} + \frac {i \omega}{\eta (r)} \bigg) G_l(r) = 0,
\end{equation}

\noindent where $\eta$ is equivalent to the magnetic diffusivity inside the planet.  It follows naturally from this equation that $G_l(r)$ has both a real and an imaginary component.

\subsubsection{Calculating The Function $G_l(r)$}

\cite{Laine2008} calculated $G_l(r)$ by splitting it up into two parts, $Y_{1,l}(r)$ and $Y_{2, l}(r)$, where $G_l(r) = Y_{1, l}(r) + i Y_{2, l}(r)$.  From here, it is straightforward to numerically estimate $G_{l}(r)$ using \texttt{scipy.integrate.solve\_bvp}, which solves first-order differential equations using the methods from \citep{Ascher1995, Kierzenka2001}.  As Equation~\ref{eqn:G_def} is a second-order differential equation, \cite{Laine2008} defined two additional functions, $Y_{3,l}(r)=Y'_{1,l}(r)$ and $Y_{4,l}(r)=Y'_{2,l}(r)$.  This enables us to describe $G_l(r)$ with a system of four first-order equations:

\begin{equation}
\begin{split}
\label{eqn:ode}
    Y'_{1,l}(r)&=Y_{3,l}(r),\\
    Y'_{2,l}(r)&=Y_{4,l}(r),\\
    Y'_{3,l}(r)& = \frac{-\omega}{\eta(r)} Y_{2,l}(r) + \frac{l(l+1)}{r^2} Y_1(r),
    \\
    Y'_{4,l}(r) &= \frac{\omega}{\eta(r)} Y_{1,l}(r) + \frac{l(l+1)}{r^2} Y_2(r),\\
\end{split}
\end{equation}

\noindent with the boundary conditions:

\begin{equation}
\begin{split}
\label{eqn:ode_bc}
    Y_{3,l}(r\approx0) - \frac{l+1}{r\approx 0} Y_{1,l}(r\approx 0) &=0, \\
    Y_{4,l}(r\approx0) - \frac{l+1}{r\approx 0} Y_{2,l}(r\approx 0) &=0, \\
    Y_{3,l}(R_p)  + \frac{1}{R_p} Y_{1,l}(R_p) - (2l+1)R_p^l & = 0,\\
    Y_{4,l}(R_p)  + \frac{1}{R_p} Y_{2,l}(R_p) & = 0.
\end{split}
\end{equation}

These equations enable us to estimate $G_l$.  However, we still cannot calculate $\phi_{inner}$ as defined in Equation~\ref{eqn:phi_def} without determining the values of $C_l^m$.  We describe the process we used for calculating $C_l^m$ in Appendix~\ref{app:clm}.

\subsubsection{Calculating the Total Power, $P_{ind}$}

Now that we have calculated the magnitude of the poloidal scalar inside the planet, we can now calculate the total power dissipation over the entire planet $P_{ind}$ induced by the changing magnetic field.  A changing magnetic field will induce an electric field $E$ and current density $J$ inside the planet, which can then dissipate energy according to the following:

\begin{equation}
    \frac{dP_{ind}}{dV} = \mathrm{Re}(J)\, \mathrm{Re}(E) = \eta  \big(\mathrm{Re}(J)\big)^2 = \frac{\eta}{\mu_0}\,\big(\mathrm{Re}(\nabla \times B)\big)^2.
\end{equation}

To determine the value of $\nabla \times B$, we follow \cite{Laine2008}.  We start with the magnetic induction equation:

\begin{equation}
\label{eqn:mag_induction}
    \frac{\partial B}{\partial t} = \nabla \times (v \times B) - \nabla \times (\eta \nabla \times B).
\end{equation}

If the magnetic diffusivity in the planet $\eta$ is high (that is, the magnetic Reynolds number is low), or the planet and star are in corotation (making $v$ small) then the second term will dominate:

\begin{equation}
\label{eqn:diffusion_approx}
    \frac{\partial B}{\partial t} \approx - \nabla \times (\eta \nabla \times B).
\end{equation}

A discussion of the validity of the diffusive limit in this scenario is included in \cite{Laine2008}, as well as in Section~\ref{ssec:rad_profile}.  If $\eta$ is low (or $v$ is high), the system may instead be in a regime in which induction dominates. \cite{Laine2012} discusses unipolar induction in super-Earths.

We also know that, when considering the poloidal of the magnetic field:

\begin{equation}
\label{eqn:b_phi}
    B = \nabla \times (\nabla \times \phi_{\hat{r}}).
\end{equation}

Plugging values in, we find 

\begin{equation}
    \nabla \times B = -\frac{i \omega }{\eta} \nabla \times \phi_{\hat{r}},
\end{equation}

\noindent and therefore

\begin{equation}
\label{eqn:pind_integral}
    P_{ind} = \frac{\omega^2}{\mu_0}\iiint \frac{1}{\eta (r)} \Bigg( \frac{1}{\sin(\theta)}\bigg(\frac{\partial \mathrm{Im}(\phi_{inner})}{\partial \varphi} \bigg)^2 + \sin(\theta)\bigg(\frac{\partial \mathrm{Im}(\phi_{inner})}{\partial \theta} \bigg)^2\Bigg) dr d\theta d\varphi,
\end{equation}

where the volume integral is over the volume of the planet.

We next need to determine the value of Im$(\phi_{inner})$.  Thankfully, we know the form of $\phi_{inner}$ and the values of the constants $C_l^m$, which allows us to write out Im$(\phi_{inner})$ in a closed-form manner:

\begin{equation}
\label{eqn:imphi_inner}
\begin{aligned}
\mathrm{Im}(\phi_{inner})=
    \mu_0\sqrt{\frac{3}{8\pi}} \sin(\theta) \bigg(
 \cos(\omega t) \cos(\varphi)\big( \mathrm{Re}(G_1)(-\nu_1^1+\nu_1^{-1}) + \mathrm{Im}(G_1)(-\mu_1^1+\mu_1^{-1}) \big)\\
+\cos(\omega t) \sin(\varphi)\big( \mathrm{Re}(G_1)(-\mu_1^1-\mu_1^{-1}) + \mathrm{Im}(G_1)(\nu_1^1+\nu_1^{-1}) \big)\\
+\sin(\omega t) \cos(\varphi)\big( \mathrm{Re}(G_1)(-\mu_1^1+\mu_1^{-1}) + \mathrm{Im}(G_1)(\nu_1^1-\nu_1^{-1}) \big)\\
+\sin(\omega t) \sin(\varphi)\big( \mathrm{Re}(G_1)(\nu_1^1+\nu_1^{-1}) + \mathrm{Im}(G_1)(\mu_1^1+\mu_1^{-1}) \big)\bigg)\\
+\mu_0\sqrt{\frac{5}{16\pi}}\big(3\cos^2(\theta)-1\big)\bigg(
 \cos(\omega t)\big( \mathrm{Re}(G_2)\nu_2^0 +\mathrm{Im}(G_2) \mu_2^0 \big) \\
+\sin(\omega t)\big( \mathrm{Re}(G_2)\mu_2^0 -\mathrm{Im}(G_2) \nu_2^0 \big) \bigg) \\
+\mu_0\sqrt{\frac{15}{32\pi}} \sin^2(\theta) \bigg(\cos(\omega t) \cos(2\varphi) \big(\mathrm{Re}(G_2)(\nu_2^2+\nu_2^{-2}) + \mathrm{Im}(G_2)(\mu_2^2+\mu_2^{-2})  \big)\\
+\cos(\omega t) \sin(2\varphi) \big(\mathrm{Re}(G_2)(\mu_2^2-\mu_2^{-2}) + \mathrm{Im}(G_2)(-\nu_2^2+\nu_2^{-2})  \big)\\
+\sin(\omega t) \cos(2\varphi) \big(\mathrm{Re}(G_2)(\mu_2^2+\mu_2^{-2}) + \mathrm{Im}(G_2)(-\nu_2^2-\nu_2^{-2})  \big)\\
+\sin(\omega t) \sin(2\varphi) \big(\mathrm{Re}(G_2)(-\nu_2^2+\nu_2^{-2}) + \mathrm{Im}(G_2)(-\mu_2^2+\mu_2^{-2})  \big)\bigg).
\end{aligned}
\end{equation}

To simplify this equation, we define some constants:

\begin{equation}
\label{eqn:A_values}
\begin{aligned}
A_{11}= \mathrm{Re}(G_1)(-\nu_1^1+\nu_1^{-1}) + \mathrm{Im}(G_1)(-\mu_1^1+\mu_1^{-1}) \\
A_{12}=  \mathrm{Re}(G_1)(-\mu_1^1-\mu_1^{-1}) + \mathrm{Im}(G_1)(\nu_1^1+\nu_1^{-1}) \\
A_{13}=  \mathrm{Re}(G_1)(-\mu_1^1+\mu_1^{-1}) + \mathrm{Im}(G_1)(\nu_1^1-\nu_1^{-1}) \\
A_{14}=  \mathrm{Re}(G_1)(\nu_1^1+\nu_1^{-1}) + \mathrm{Im}(G_1)(\mu_1^1+\mu_1^{-1}) \\
A_{15}=  \mathrm{Re}(G_2)\nu_2^0 +\mathrm{Im}(G_2) \mu_2^0  \\
A_{16}= \mathrm{Re}(G_2)\mu_2^0 -\mathrm{Im}(G_2) \nu_2^0 \\
A_{17}= \mathrm{Re}(G_2)(\nu_2^2+\nu_2^{-2}) + \mathrm{Im}(G_2)(\mu_2^2+\mu_2^{-2})  \\
A_{18}= \mathrm{Re}(G_2)(\mu_2^2-\mu_2^{-2}) + \mathrm{Im}(G_2)(-\nu_2^2+\nu_2^{-2})  \\
A_{19}= \mathrm{Re}(G_2)(\mu_2^2+\mu_2^{-2}) + \mathrm{Im}(G_2)(-\nu_2^2-\nu_2^{-2})  \\
A_{110}= \mathrm{Re}(G_2)(-\nu_2^2+\nu_2^{-2}) + \mathrm{Im}(G_2)(-\mu_2^2+\mu_2^{-2}),
\end{aligned}
\end{equation}

\noindent so that the form of $\phi_{inner}$ is:

\begin{equation}
\begin{aligned}
  \mathrm{Im}(\phi_{inner})= \mu_0 \sqrt{\frac{3}{8\pi}} \sin(\theta) \bigg(
 \cos(\omega t) \cos(\varphi)A_{11}+\cos(\omega t) \sin(\varphi)A_{12}+\sin(\omega t) \cos(\varphi)A_{13}+\sin(\omega t) \sin(\varphi)A_{14}\bigg)\\
+\mu_0\sqrt{\frac{5}{16\pi}}\big(3\cos^2(\theta)-1\big)\bigg(\cos(\omega t)A_{15} +\sin(\omega t)A_{16} \bigg) \\
+\mu_0\sqrt{\frac{15}{32\pi}} \sin^2(\theta) \bigg(\cos(\omega t) \cos(2\varphi) A_{17}+\cos(\omega t) \sin(2\varphi) A_{18}+\sin(\omega t) \cos(2\varphi) A_{19}+\sin(\omega t) \sin(2\varphi) A_{110}\bigg).
\end{aligned}
\end{equation}

We then can insert this into Equation~\ref{eqn:pind_integral} and get:

\begin{equation}
\label{eqn:pind_circ}
\begin{aligned}
    P_{ind} =  \int \frac{\omega^2 \mu_0}{\eta (r)}  \Bigg( \cos^2(\omega t)\bigg(A_{11}^2+A_{12}^2 + 3\big(A_{17}^2  + A_{18}^2\big) + 6 A_{15}^2 \bigg) + \sin^2(\omega t) \bigg(A_{13}^2+A_{14}^2 + 3\big(A_{19}^2 + A_{110}^2\big) + 6 A_{16}^2 \bigg) \\
    + 2 \cos(\omega t) \sin(\omega t) \bigg(A_{11} A_{13} + A_{12} A_{14} + 3\big( A_{17}A_{19} + A_{18}A_{110}\big) + 6 A_{15}A_{16} \bigg) \Bigg)dr.
\end{aligned}
\end{equation}

This is analogous to Equation~13 in \cite{Laine2008}, and is similar to their results within less than an order of magnitude.  %It is possible that our results differ slightly due to some small error in calculation in some intermediate steps.

In summary, we can calculate the power dissipated inside of a planet with the following steps:

\begin{enumerate}
    \item Set a value for the planet radius $R_p$, the planet's orbital period $p_{orb}$, the star's rotation period $p_{rot}$, the stellar magnetic dipole $m$, the angle of the star's spin axis $\alpha$, and the orbital distance $D$.
    \item Use $p_{orb}$ and $p_{rot}$ to calculate the angular frequency $\omega = 2\pi/p_{rot} - 2\pi/p_{orb}$.
    \item Select a model for the magnetic diffusivity inside the planet, $\eta(r)$.
    \item Use $R_p$, $\eta(r)$, and $\omega$ to calculate $G_l(r)$ by inserting Equations~\ref{eqn:ode} and \ref{eqn:ode_bc} into a numerical differential equation solver.
    \item Use $G_l(r)$ to calculate the values of the coefficients $C_l^m$ in Equation~\ref{eqn:phi_def} by evaluating $\phi_{inner} = \phi_p + \phi_\star$ at the radius of the planet.
    \item Use $G_l(r)$, in addition to the $C_l^m$ values, to calculate the $A$ coefficients (Equation~\ref{eqn:A_values}).
    \item Integrate the power density over the volume of the planet to find the total power $P_{ind}$ using Equation~\ref{eqn:pind_circ}.
\end{enumerate}

\subsection{The Magnetic Diffusivity, $\eta$}
\label{ssec:diffusivity}
While most of the variables used to calculate $P$ are very straightforward and can be derived directly from observations, the magnetic diffusivity $\eta(r)$ is a function of the internal conditions of the planet and is thus likely substantially more complex. 

If the dominant source of resistivity in the planetary atmosphere is the result of electron-neutral collisions, we can use the equation for conductivity $\sigma$ from \cite{Blaes1994}:

\begin{equation}
    \frac{c^2}{4\pi \sigma} =  230\bigg(\frac{n_n}{n_e}\bigg) \sqrt{T} \,\mathrm{cm}^2 \mathrm{s}^{-1}
\end{equation}

\noindent where $T$ is the temperature, $n_e$ is the number density of electrons, and $n_n$ is the number density of neutral atoms, and find (after converting to SI units) the diffusivity $\eta = \sigma^{-1} \mu_0^{-1}$ to be:

\begin{equation}
\label{eqn:eta_bb}
    \eta = 0.023 \bigg(\frac{n_n}{n_e}\bigg) \sqrt{T}\,\mathrm{m}^2 \mathrm{s}^{-1}.
\end{equation}

\begin{figure}
    \centering
    \includegraphics[width=0.5\linewidth]{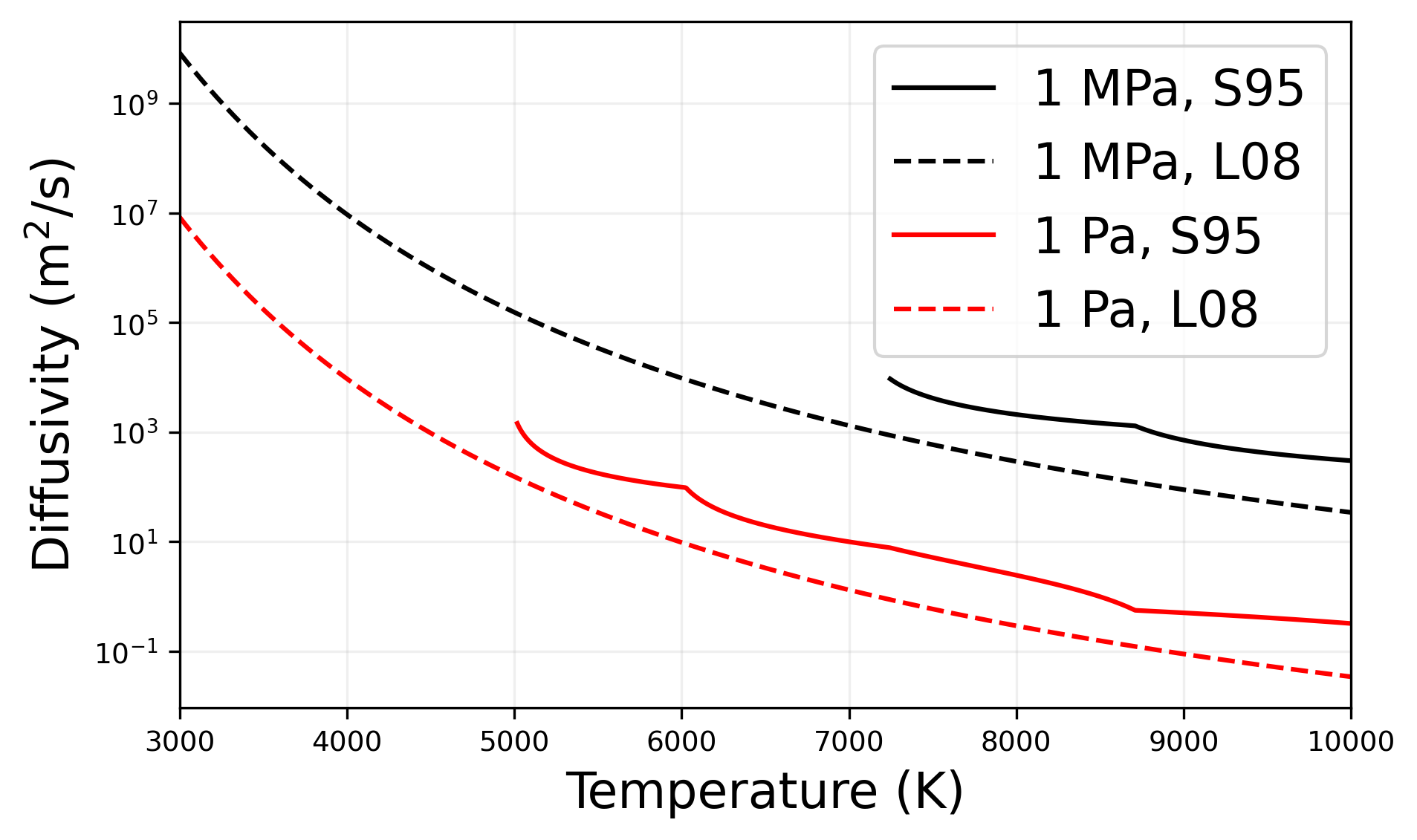}
    \caption{The magnetic diffusivity vs. temperature of a hydrogen-helium atmosphere at various pressures.  We show the diffusivity as inferred by the \protect\cite{Saumon1995} equation-of-state tables as solid lines and the diffusivity from the simplified \protect\cite{Laine2008} formula as dashed lines.}
    \label{fig:ne_nn}
\end{figure}

Unfortunately, $n_e/n_n$ is a complicated function of temperature that is difficult to write as a closed-form solution, as there are many different potential sources of electrons in a planet atmosphere.  For an initial estimate, we used the equation of state tables from \cite{Saumon1995} (as reproduced in the MESA code) for a hydrogen-helium atmosphere to estimate $n_e/n_n$ as a function of temperature and pressure, following the methods used in \cite{Chang2012}.  We show this $\eta$ in Figure~\ref{fig:ne_nn}.

Our calculated value for $\eta$ appears to be ten times higher than that from \cite{Laine2008}:

\begin{equation}
\label{eqn:eta_l08}
    \eta = 1.28 *10^{-2} \frac{\sqrt{P(r)}}{T(r)^{3/4}} e^{78909/T}\,\mathrm{m}^2 \mathrm{s}^{-1},
\end{equation}

\noindent which was estimated using the Saha equation and assuming that the ionization fraction in the region of Ohmic dissipation is small, which makes sense given the fact that we anticipate that the majority of the magnetic field is dissipated near the surface (see Section~\ref{ssec:rad_profile}), where temperatures are relatively low.  However, we note that the \cite{Laine2008} equation assumes a conductivity ten times higher than what they measured from their calculations, in an effort to account for additional (non-hydrogen) atoms in the planetary atmosphere.

In our base model, we will use the equation for $\eta$ from \cite{Laine2008}, which is valid over a larger range of pressures and temperatures than our interpolations over the \cite{Saumon1995} tables.  We note that, as shown in \cite{Laine2008}, the final power induced by Ohmic dissipation is only weakly dependent on the value of $\eta$ (where a change in $\eta$ by several orders of magnitude only affects $P$ by a factor of a few), so it is acceptable to use an approximation in this scenario. 

However, to explore the influence of the shape of the $\eta$ curve on our results, we will also consider a second $\eta$ model inspired by the fact that metal species, if present in the planetary envelope, are much more easily ionized than hydrogen or helium.  At temperatures lower than the ionization temperature of hydrogen (which will be relevant in the planet's upper atmosphere), these metal species will likely be the dominant source of electrons, therefore reducing the diffusivity described in Equation~\ref{eqn:eta_bb}.  Inspired by the work in \cite{Laine2008}, we considered the outer envelope of the planet to be made up of atomic hydrogen, helium, sodium, and potassium, with sodium and potassium being the dominant sources of electrons (given their relatively low ionization energies).  We assumed that the various elements had relative abundances corresponding with those in the solar photosphere, with values taken from \cite{Asplund2009}.  This is a simplification of the real conditions in the atmosphere of a planet (which may contain molecular hydrogen, as well as numerous other elements that could contribute electrons), but will at least provide a rough estimate of the conditions in the upper atmosphere.  We then solved the Saha equation for sodium and potassium at the values of $T$ and $\rho$ dictated by our models in Section~\ref{ssec:pt}, assuming that hydrogen and helium remain neutral at the relevant probed temperatures.  This enabled us to estimate $n_n/n_e$ in the upper atmosphere, creating a unique model for $\eta$ that could capture the conditions in the upper planetary atmosphere.  We also created a model that implemented this version of $\eta$ in the upper atmosphere and transitioned into the Equation~\ref{eqn:eta_l08} in the deep atmosphere, which should theoretically capture both the electrons from alkali ionization in the upper atmosphere and the additional electrons from hydrogen and helium in the deep atmosphere, which would be the dominant sources of electrons in those respective zones.  We discuss the effects of our choice of $\eta$ on the power generation in Section~\ref{sec:pind_mag}.

\subsection{The Planet's Temperature-Pressure Profile}
\label{ssec:pt}

As shown in Section~\ref{ssec:diffusivity}, the diffusivity of a planet is dependent both upon its temperature and pressure.  Therefore, it is necessary to estimate the temperature-pressure profile of a planet in order to determine the functional form of $\eta$, which in turn allows us to calculate $P$.  

To calculate the temperature-pressure profile of the planet, we used \texttt{GASTLI} \citep{Acuna21, Acuna24}, which enables us to estimate the temperature and pressure profiles of the planet assuming a water/rock core coupled to a simple envelope composed of H/He and water. A majority H/He atmosphere is a reasonable assumption in our calculations, assuming that we are studying Jupiter-like giant planets with Jupiter-like envelopes.   As discussed in \cite{Laine2008} and derived later in this paper, the majority of the energy dissipation occurs in the outer envelope of the planet, making the exact nature of the planet's core make-up unimportant in our calculations.  

\texttt{GASTLI} is applicable to planets with masses between roughly $0.05-6$\,$M_J$ and $T_{eq}<1000$\,K, though it can extrapolate its models to parameters outside of those ranges.  While we are interested in studying short-period planets with higher irradiation temperatures than $1000$\,K, we found that \texttt{GASTLI} appears to converge to reasonable values (in which the surface temperature is close to the irradiation temperature) even for very high values of $T_{eq}$.  The only issue we found is that there were occasionally small discontinuities in the profiles at the interface between the interior and atmosphere \texttt{GASTLI} models, which we resolved using the \texttt{CubicSpline} interpolator from \texttt{scipy}.  We also note (as discussed in Section~\ref{ssec:irradiation_comp}) that the bulk power dissipation in a planet is relatively insensitive to the values assumed for $\eta$, which is derived from the temperature-pressure profile.  Therefore, we do not anticipate that small inaccuracies in the profile will heavily influence our results.

To estimate the profile with \texttt{GASTLI}, the user must input a mass, internal temperature, equilibrium temperature, core-mass-fraction (CMF), envelope metallicity, and C/O ratio.  Unless otherwise stated, we assumed an internal temperature of 100\,K, a CMF of 0.01, and a solar metallicity and C/O ratio.  For each model, we adjusted the equilibrium temperature to be the value based on the star's characteristics and the planet's semi-major axis, assuming full heat redistribution and a zero albedo \citep[most observed hot Jupiters appear to have low albedos, see, e.g.,][]{Heng2013, Krenn2023}.  We also typically assumed a planet mass of $1\,M_J$, and used \texttt{GASTLI} to infer the planet's radius given these input parameters.

\section{The Magnitude of $P_{ind}$}
\label{sec:pind_mag}

\subsection{Radial Profile of Power Dissipation within the Planet}
\label{ssec:rad_profile}

\begin{figure}
    \centering
    \includegraphics[width=1\linewidth]{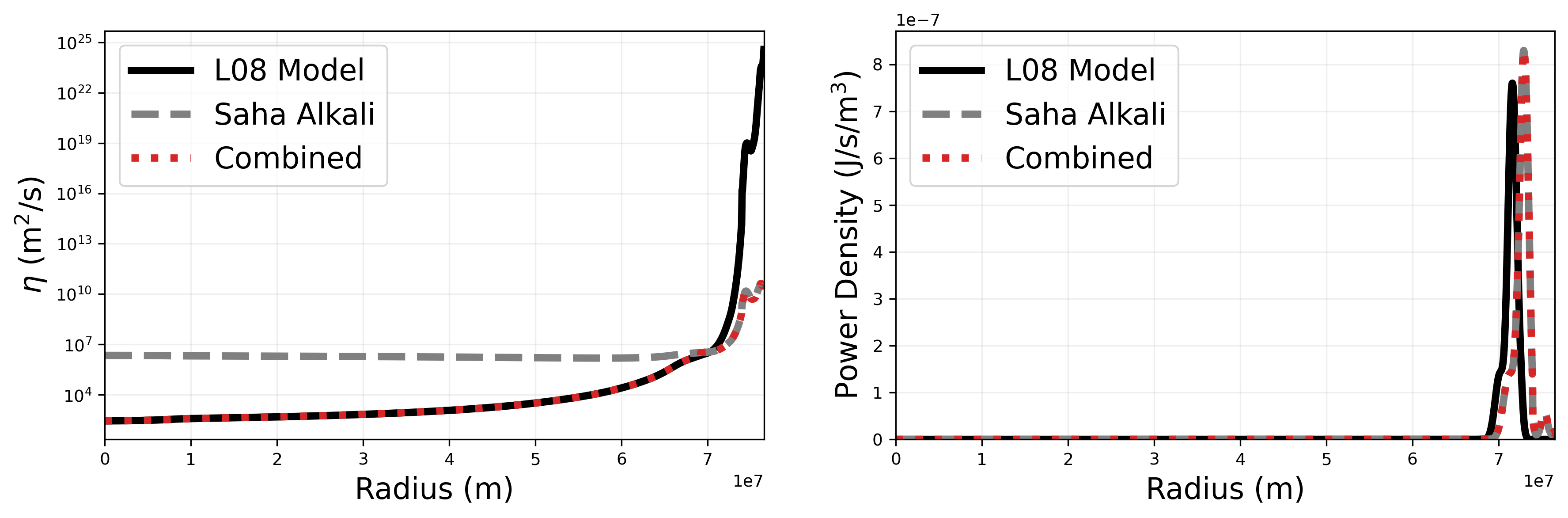}
    \caption{The magnetic diffusivity (left panel) and power dissipation (right panel) within a $M_p=1$\,$M_J$ planet with a 3d orbital period orbiting a star with a rotation period of 5d and a surface magnetic field of 1 kG as a function of radius, where $R=0$ is the center of the planet.  The solid lines denote the model where we used Equation~\ref{eqn:eta_l08} to estimate the value of $\eta$, the dashed lines denote the model in which we used the Saha equation for potassium and sodium, and the dotted lines denote our model that uses the Equation~\ref{eqn:eta_l08} model in the interior and the alkali model in the envelope.  It is clear that $\eta$ increases dramatically towards the exterior of the planet, and the vast majority of the inductive power is deposited near the top of the atmosphere.}
    \label{fig:dissipation_location}
\end{figure}

With our model, we can calculate $\mathcal{P}$, where $P_{int} = \int \mathcal{P} r^2 dr$.   This enables us to determine where, exactly, power is deposited in the planet as a function of radius.  Using our various models of $\eta$ from Section~\ref{ssec:diffusivity} and a \texttt{GASTLI} temperature-pressure model for a 1\,M$_J$ planet evaluated over a grid of 100,000 evenly-spaced points across the radius of the planet, we find that the vast majority of power dissipation occurs near the surface of the planet, as is shown in Figure~\ref{fig:dissipation_location}.

The power dissipation is insensitive to the value of $\eta$ in the deep planet interior.  We found no significant difference in $P_{ind}$ between a model that ignored the effects of hydrogen/helium ionization in the planetary interior versus a model that included these effects (through Equation~\ref{eqn:eta_l08}), despite the fact that the actual value of $\eta$ in these two models differs by several orders of magnitude. This means that our results are likely somewhat independent of the behavior of the interior of the planet.  This is useful, as it means that our results can likely be generalized to any planet with a thick gaseous envelope, without having to be mindful of the planet's core composition.

\begin{figure}
    \centering
    \includegraphics[width=0.5\linewidth]{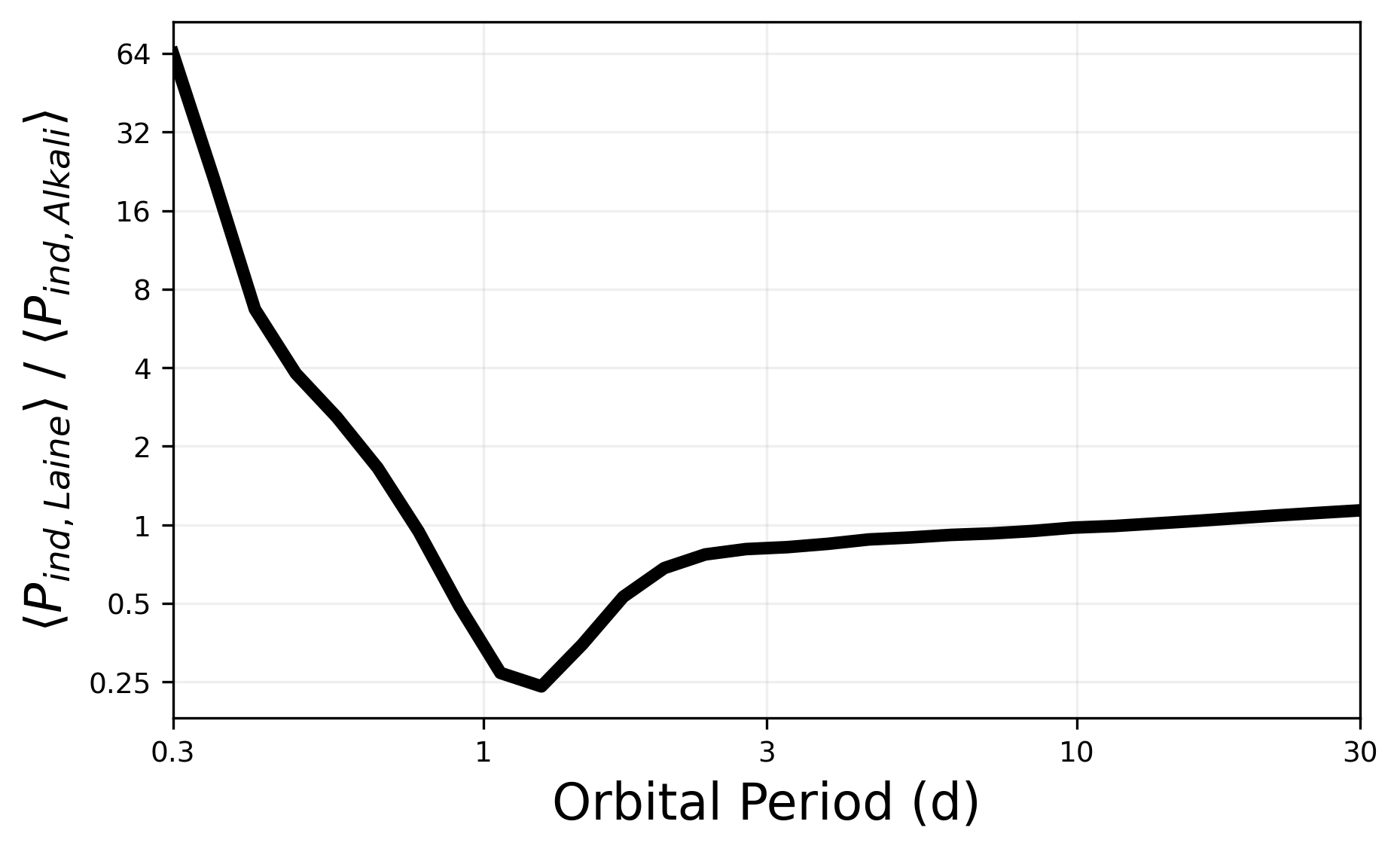}
    \caption{The comparison between $P_{ind}$ for a planet with the two different $\eta$ prescriptions as a function of orbital period for a $M_p=1$\,$M_J$ planet orbiting a star with a rotation period of 5d and a surface magnetic field of 1 kG.  Overall, it is clear that our choice of $\eta$ prescription is only significant for planets with periods below about 2d.}
    \label{fig:pind_eta_prescription}
\end{figure}

In addition, while our prescription for $\eta$ in the outer envelope does appear to change the value of $P_{ind}$, it only causes significant changes in models where $p_{orb}$ is very short.  In our default model in Figure~\ref{fig:dissipation_location}, including the alkali metals only increased the total induced power around 20\%, even though (once again) the actual value of $\eta$ differed from the Equation~\ref{eqn:eta_l08} model by many orders of magnitude.  We show how our $\eta$ prescription changes $P_{ind}$ as a function of orbital period in Figure~\ref{fig:pind_eta_prescription}.  In general, the change appears to be less than one order of magnitude for orbital periods where $p_{orb} > 2$d.  For extremely short orbital periods, the \cite{Laine2008} model can result in order-of-magnitude higher values of $P_{ind}$- this may be due to the fact that, at very high temperatures, hydrogen and helium can ionize even in the upper atmosphere, resulting in additional free electrons that the alkali model does not account for.  In the interest of producing a conservative upper limit on the total induced power from magnetic induction, we will proceed with using the \cite{Laine2008} $\eta$ model to calculate $P_{ind}$ in later sections.  However, we note that, for orbital periods longer than $2$d, our choice of model makes very little difference.

Our calculated $\eta$ profiles enable us to determine the validity of the assumption we made in Equation~\ref{eqn:diffusion_approx}. In general, the magnetic Reynolds number $R_m$ describes the ratio between the strength of the inductive and diffusive terms in the magnetic induction equation.  Therefore, if diffusion dominates, $R_m<<1$.   $R_m$ is defined as:

\begin{equation}
    R_m = \frac{v l}{\eta}
\end{equation}

where $v$ is the velocity and $l$ is the length scale.  The relevant velocity in this scenario is the wind speed of the planet.  As a point of reference, \cite{Seidel2026} studied the wind speeds of seven hot Jupiters and found wind speeds ranged from around 1 --- 7\,km$s^{-1}$, so we will quote a wind speed of 7\,km$s^{-1}$ to attempt to calculate an upper limit on $R_m$.  $l$ is related to the depth of magnetic field penetration, and therefore is, at the very highest, $R_p$, though we show in Figure~\ref{fig:dissipation_location} that the magnetic field may only penetrate through the outer envelope of the planet.  This results in a Reynolds number of:

\begin{equation}
    R_m \approx \frac{v l}{\eta} \approx  \bigg(\frac{v_{wind}}{7~\mathrm{km}\mathrm{s}^{-1}}\bigg)  \bigg(\frac{l}{~0.1R_{J}}\bigg) \bigg(\frac{\eta}{4.8 * 10^{10}}\bigg)^{-1}.
\end{equation}

As shown in Figure~\ref{fig:dissipation_location}, in our nominal model, $\eta$ can take on a broad range of values, from below $10^4$\,m$^2$s$^{-1}$ in the deep interior to $10^{25}$\,m$^2$s$^{-1}$ near the edge of the envelope.  Therefore, it is reasonable to assume that diffusion is dominant in the outer envelope of the planet, where $\eta$ is very high ($\eta>10^{12}$m$^2$s$^{-1}$), especially in scenarios where the magnetic field penetration is shallow or the wind speed is low.  However, induction is far more important in the planet's deep interior, where the first term in Equation~\ref{eqn:mag_induction} dominates.  We refer the reader to \cite{Laine2012}, which discusses the nature of the magnetic field properties in this scenario in high-conductivity super-Earths.  Our math shows us that it is possible that the inductive term is relevant to the field dissipation in the intermediate layers of hot Jupiters, but we leave a more concrete mathematical study of it to future works.  

Our alkali-$\eta$ model is far less constraining in terms of the Reynolds number, as it results in lower values of $\eta$ in the planetary envelope. $R_m$ can still be less than one if the planetary wind speeds are low (low $v$) or the planet has a long ($>3$d) orbital period (resulting in fewer free electrons and therefore a higher $\eta$), but we cannot guarantee that $R_m$ is low in all hot Jupiters.  Notably, this means that the diffusive atmosphere assumption may not necessarily be accurate for windy, short-period hot Jupiters- such planets may instead be dominated by magnetic induction.  A significantly more complex model of atmospheric diffusivity (taking into account, for example photoionization) may be useful for determining the true nature of diffusion versus induction in the outer atmospheres of giant planets, but is beyond the scope of this current work.

We also examined the relative amount of power dissipation that occurs due to the $l=1$ and $l=2$ harmonics of the stellar magnetic field by performing our calculations in both the nominal scenario and a scenario in which the second-order terms in Equation~\ref{eqn:psi_grouped} are equal to zero.  As implied by the scaling in Equation~\ref{eqn:psi_grouped}, the $l=2$ terms are suppressed by a factor of $r/D$ relative to the $l=1$ terms, meaning that the $l=2$ terms are more likely to be meaningful for planets on short orbits.  However, we found that, for orbital periods greater than 0.5d, the change in induced  power at any point within the radial profile of the planet changed by less than 1\%.  Even for a hypothetical planet with a (likely unphysically short) orbital period of 0.1d, omitting the $l=2$ term only affects the induced power within the planet by around 10-20\%.

\subsection{How $P_{ind}$ Scales With System Parameters}

There are many input parameters that are relevant to the magnitude of $P_{ind}$.  We discuss these parameters, as well as their scaling relationships with $P_{ind}$, in this this section.

\subsubsection{The Magnetic Dipole Moment, $m$}

One of the most obvious parameters is the value of the magnetic dipole moment, $m$.  $m$ is an input into the value of the magnetic field outside of the planet (Equation~\ref{eqn:phi_star}) and sets the magnitude of the magnetic field inside the planet (through the $C_l^m$ coefficients in Equation~\ref{eqn:phi_def}).  Therefore, a higher value of $m$ will result in a larger amount of dissipated power.  $m$ has a value of $2 \pi B_{surf} R_\star^3 / \mu_0$, where $B_{surf}$ is the surface magnetic field of the star and $R_\star$ is the star's radius. 

Let us consider what are reasonable values for $m$, as this will allow us to determine the typical magnitude of the inductive heating term.  In order to maximize the potential strength of this term, we are interested in systems where $m$ is maximized.

The Sun, which is an excellent point of reference for old, main-sequence stars, has a surface magnetic field of roughly 1\,G \citep[e.g.,][]{Babcock1961}.  This results in an estimated magnetic dipole moment of roughly $10^{29}$\,A\,m$^2$.  Young sun-like stars, meanwhile, as shown by \cite{Kochukhov2020}, tend to have higher surface magnetic fields ($\sim 1$\,kG), resulting in higher magnetic dipole moments ($\sim 10^{32}$\,A\,m$^2$). As sun-like stars age, their magnetic field strengths tend to drop \citep{Folsom2016}.

Meanwhile, M dwarfs, which are more convective, tend to host stronger magnetic fields.  \cite{Wanderley2024} studied a sample of 29 planet-hosting M dwarfs and found that they had typical magnetic field strengths on the order of 0.1-1\,kG, while \cite{Reiners2022}, in a sample of 292 M dwarfs, found a M dwarf with a magnetic field of 8\,kG.  Studying the \cite{Reiners2022} sample, the highest M dwarf dipole moments are on the order of 10$^{32}$\,A\,m$^2$, which are comparable to those of young sun-like stars.

As young stars tend to have stronger magnetic fields, we next consider T Tauri stars.  As our proxy for T Tauri stars, we consider TW Hya, which was found in \cite{Donati2024} to have a typical surface magnetic field strength of 3.6\,kG \citep[which appears to be consistent with the magnetic fields of other T Tauri stars, see, e.g.,][]{Lavail2019}.  This, combined with its radius of $1.16\,\pm\,0.13\,R_\odot$, lends it a dipole moment of $9*10^{32}$\,A\,m$^2$.  The observed magnetic fields of T Tauri stars (which appear to be on the order of a few kilogauss) agree with the theoretical models of \cite{Ahmad2025}, which indicated that low-mass stars will typically form with comparable magnetic fields. 

We also consider giant stars.  As $m\propto R_\star^3$, evolved giant stars may superficially appear to have the potential to be the highest-$m$ sources.  However, giant stars tend to have lower surface magnetic fields- \cite{Auriere2015} found that all of the studied giant stars in their sample of G-K giants had $B_{surf}<100$\,G.  In their sample, the majority of giants with superlative magnetic moments are those with $R_\star>10\,R_\odot$, where the magnetic dipole moment can be as high as $6*10^{34}$\,A\,m$^2$.  However, it can be very difficult to detect planets around stars this large given their small RV and transit signals.

Another category of stars holds promise for having extremely powerful dipole fields.  mCP stars (magnetically chemically-peculiar stars) are a population of intermediate-mass stars which can host dipole magnetic fields as high as 20-30 kG \citep{Sikora2019}. The fields on these stars are are believed to be fossil remnants of interstellar magnetic fields that were present during the initial protostar formation \citep[see, e.g.,][]{Donati2009}.  The highest dipole magnetic field in \citep{Sikora2019} (which performs a volume-limited survey of mCP stars within 100 pc) belongs to HD\,65339, which has a dipole magnetic field of roughly 18.1\,kG and a radius of around 2.3\,$R_\odot$ \citep{Landstreet1988}.  This would result in a dipole magnetic moment of roughly $4*10^{34}$\,A\,m$^2$.  While this magnetic moment is very high, we have yet to discover any exoplanets orbiting these types of stars.  We therefore suggest caution in adopting mCP-like values when estimating $P$. 

White dwarfs may also possess fairly high surface magnetic fields.  As discussed in \citep{Ferrario2015}, magnetic fields on magnetic white dwarfs can vary from 10$^3$ --- 10$^9$ G, which would imply magnetic moments as high as $10^{33}$\,A\,m$^2$. However, magnetic white dwarfs only make up a minority of all observed white dwarfs \cite{Sion2014}, meaning that the vast majority of white dwarfs have substantially smaller dipole moments.

Finally, we consider magnetars, which would have some of the highest surface magnetic field strengths of all planet hosts.  \cite{Gomes2019} calculated the magnetic fields necessary to distort a neutron star, and found an upper limit to the magnetic moment of $2*10^{31}$\,A\,m$^2$, which would correspond to a surface magnetic field of $10^{16}$\,G, which is comparable to the magnetic fields observed in some of the highest-field neutron stars.

\begin{figure}
    \centering
    \includegraphics[width=0.5\linewidth]{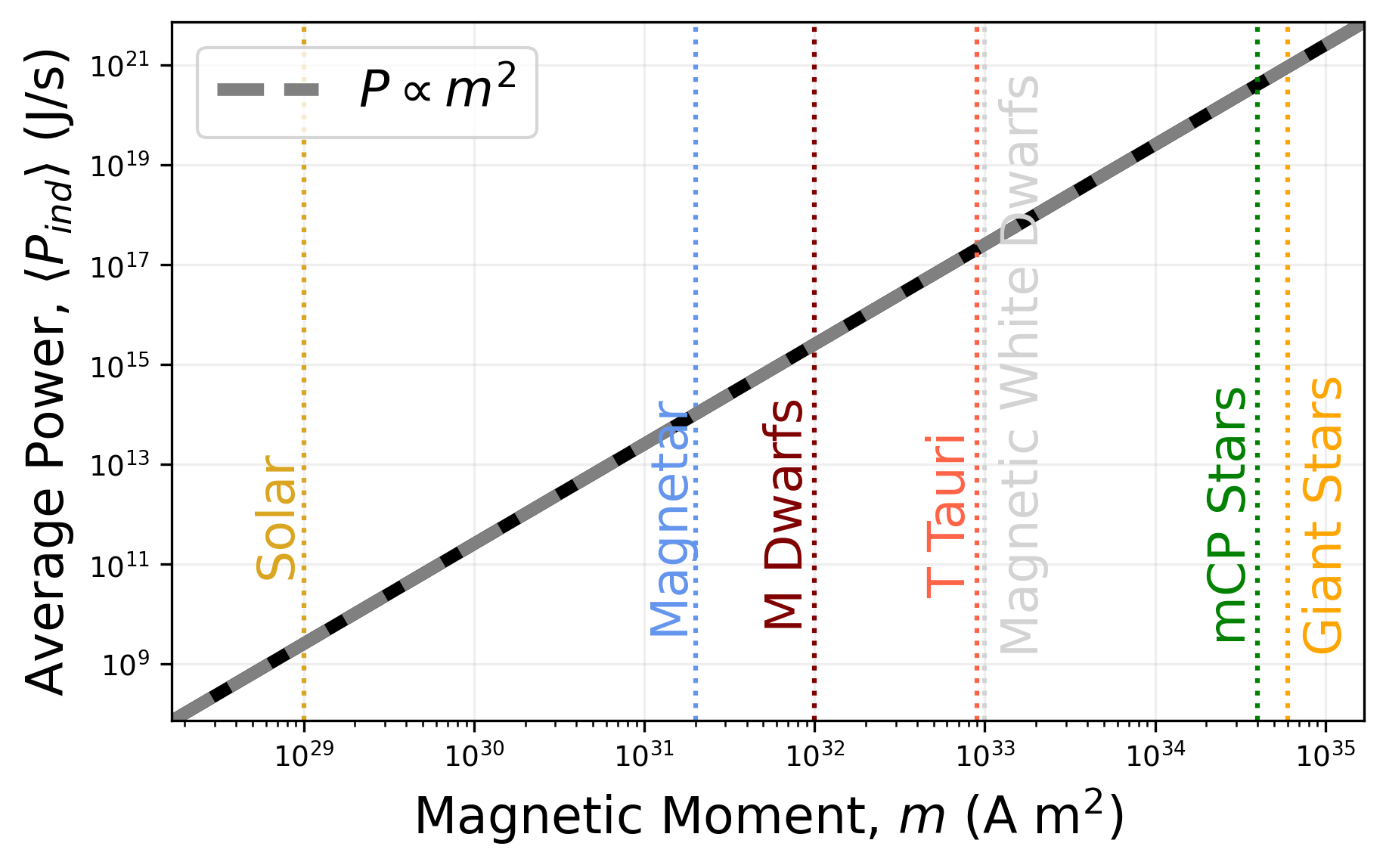}
    \caption{The average power dissipation within a $M_p=1$\,$M_J$ planet with a 3d orbital period as a function of the magnetic dipole moment of the host star, which is sun-like and has a rotation period of 5d.  We include, as vertical lines, the dipole moments of various types of stars.  We note that most of these lines (e.g., for giant stars, M dwarfs, white dwarfs, and magnetars) represent upper limits based on our current observations, and that most stars in these individual classes likely have lower dipole moments.}
    \label{fig:m_power}
\end{figure}

In Figure~\ref{fig:m_power}, we show the power dissipation within the planet as a function of the magnetic moment, with various different values for reference stars highlighted.  We observe that $P\propto m^2$, which makes sense as $\phi_\star\propto m \propto \phi$ and $P\propto \phi^2$.

\subsubsection{$\alpha$}

\begin{figure}
    \centering
    \includegraphics[width=0.5\linewidth]{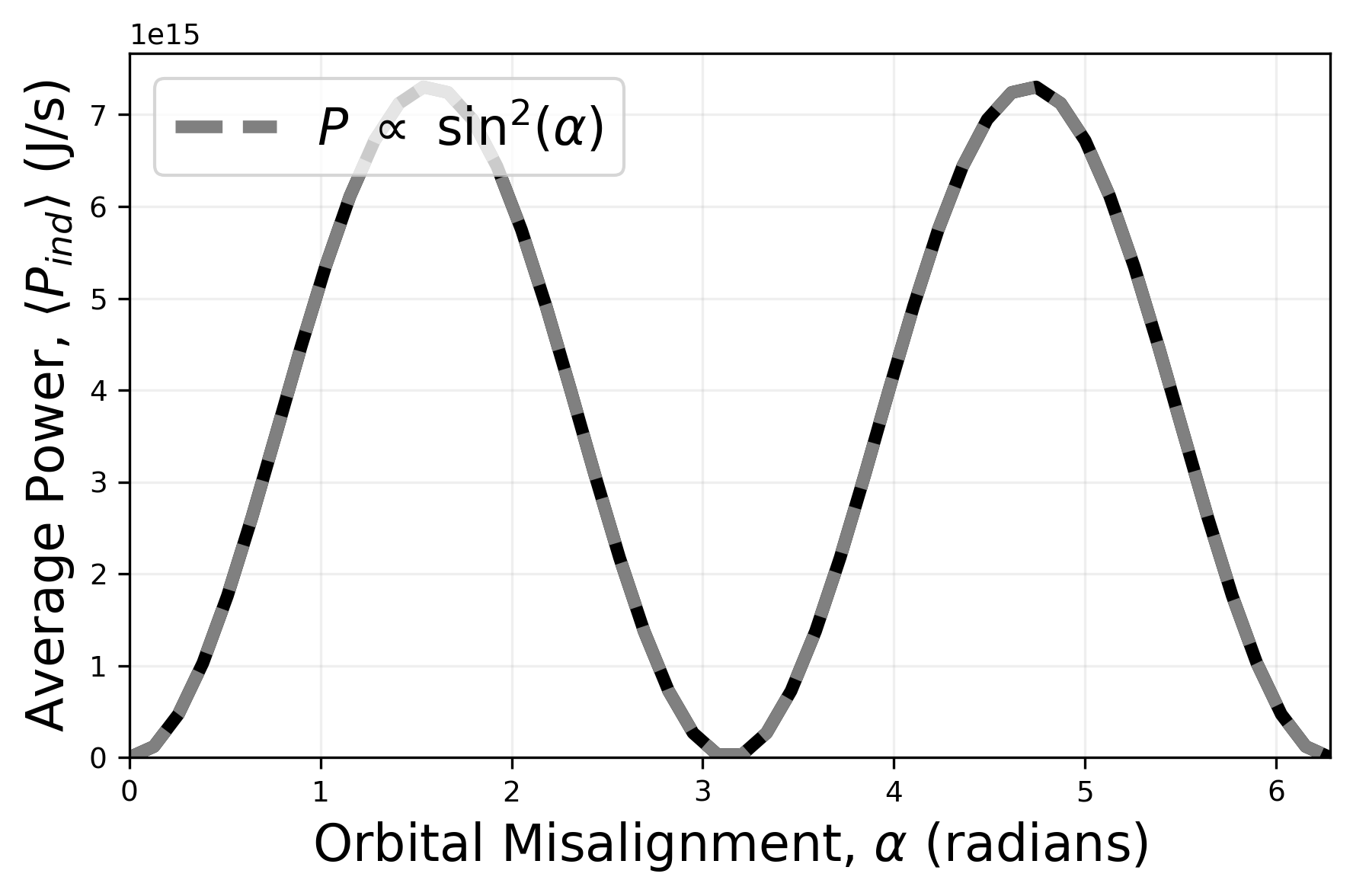}
    \caption{The average power dissipation within a $M_p=1$\,$M_J$ planet with a 3d orbital period orbiting a sun-like star with $B_{surf}=1$\,kG and $p_{rot}=$5d as a function of $\alpha$.}
    \label{fig:alpha_power}
\end{figure}

$\alpha$ describes the angle between the stellar rotation axis and magnetic dipole.  If $\alpha$ is equal to zero, the star's dipole is aligned with its rotation axis, and, if $\alpha$ is equal to $\pi/2$, they are perpendicular.  Similarly to the case with $m$, the scaling between $\alpha$ and $P$ is straightforward to derive by examining Equation~\ref{eqn:phi_star}.  As $\phi_\star\propto \sin(\alpha)$, we anticipate that $P\propto \sin^2(\alpha)$- a result that we recover with our numerical calculations (see Figure~\ref{fig:alpha_power}).   We note that this relationship becomes more complicated if the planet has an eccentric orbit, which we will discuss in Section~\ref{ssec:ecc}.

\subsubsection{$p_{orb}$ and $p_{rot}$}

The planet's orbital period $p_{orb}$ is relevant for several reasons.  Firstly, it influences the distance $d$ in Equation~\ref{eqn:phi_star}, meaning that a planet that has a longer orbital period is further away from its star and therefore undergoes less inductive heating (as $\phi_\star\propto d^{-3}$).  

In addition, the planet's period also influences $\omega$, which we can write as:

\begin{equation}
\label{eqn:omega_def}
    \omega = \frac{2\pi}{p_{rot}} -\frac{2\pi}{p_{orb}} = \frac{2\pi(p_{orb}-p_{rot})}{p_{orb} p_{rot}}.
\end{equation}

\begin{figure*}
    \centering
    \includegraphics[width=1\linewidth]{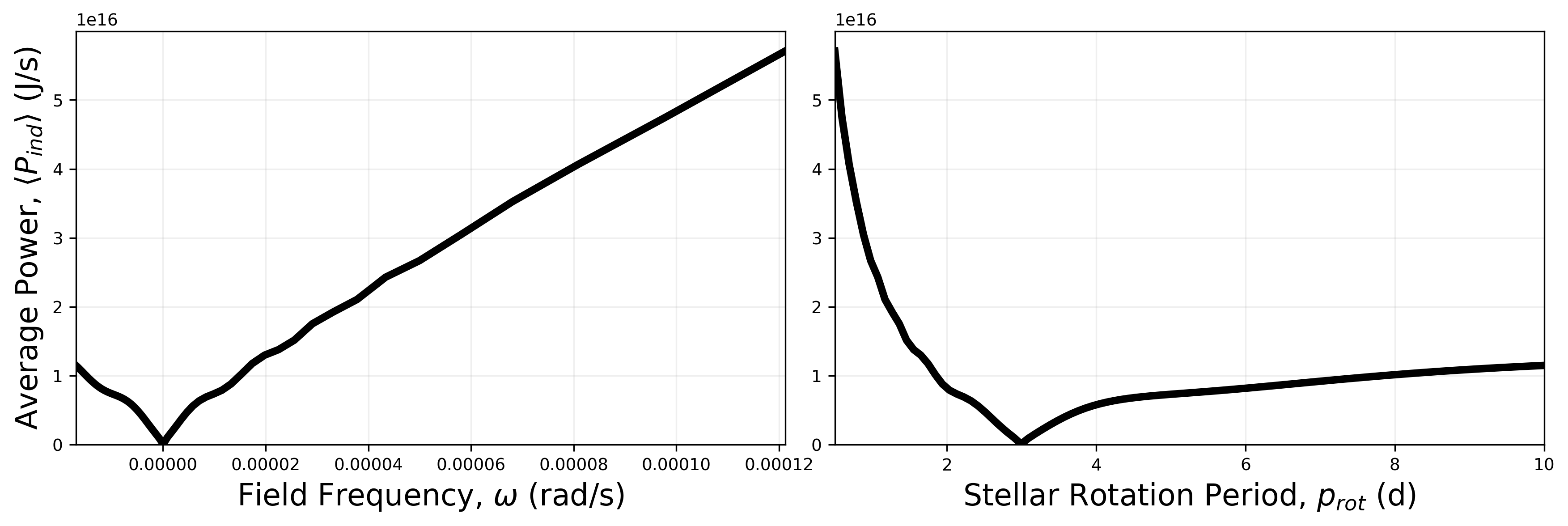}
    \caption{The average power dissipation within a $M_p=1$\,$M_J$ planet with a 3d orbital period orbiting a sun-like star with a surface magnetic field of $B_{surf}=1$\,kG.  The left panel shows the amount of power dissipated as a function of $\omega$, and the right panel shows the dissipation as a function of $p_{rot}$.}
    \label{fig:omega_power}
\end{figure*}

Interestingly, if $p_{rot} = p_{orb}$, $\omega$ is exactly equal to zero, there is no time-dependent component of the magnetic field and therefore no dissipated power.  Conversely, as shown in Figure~\ref{fig:omega_power}, higher values of $|\omega|$ result in a larger amount of dissipated power $P$.   We maximize $|\omega|$ at a fixed $p_{rot}$ when $p_{orb}$ is low, meaning that a shorter orbital period also increases $P$ through its impact on $\omega$.

The stellar rotation period $p_{rot}$ behaves similarly- as it is in the denominator in \ref{eqn:omega_def}, more rapidly rotating stars will typically result in a greater amount of energy dissipation (unless $p_{rot}=p_{orb}$).

\begin{figure}
    \centering
    \includegraphics[width=0.5\linewidth]{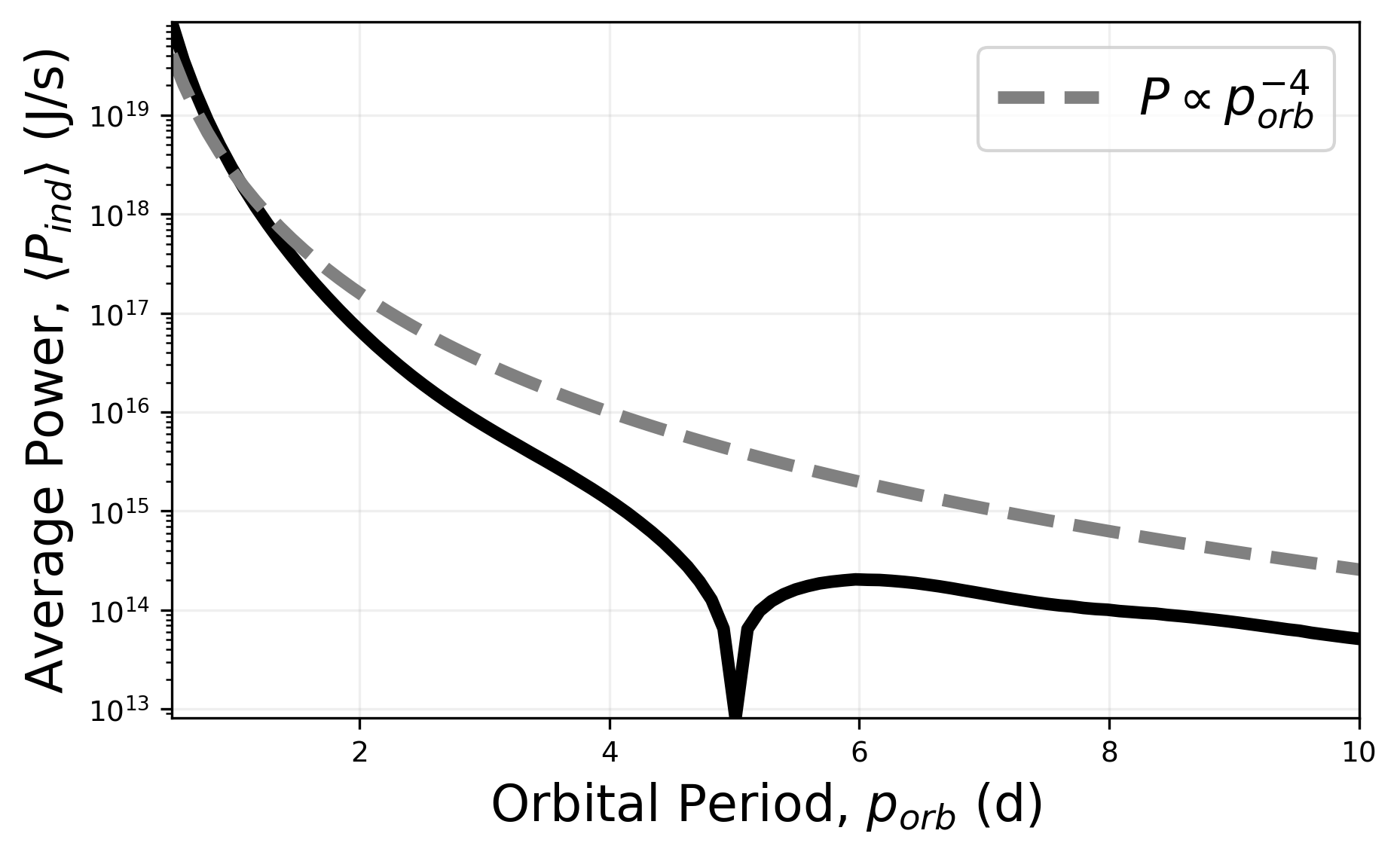}
    \caption{The average power dissipation within a $M_p=1$\,$M_J$ planet orbiting a sun-like star with a rotation period of 5d and a surface magnetic field of $B_{surf}=1$\,kG as a function of its orbital period.  We also overplot a gray dashed line corresponding to $P\propto p_{orb}^{-4}$ to highlight the anticipated scaling relationship.}
    \label{fig:porb_power}
\end{figure}

In Figure~\ref{fig:porb_power}, we show the effect of $p_{orb}$ on the power dissipation inside of the planet. In general, a shorter orbital period results in a dramatic increase in the power dissipation as roughly $P\propto p_{orb}^{-4}$.  This scaling makes sense, as $d\propto p_{orb}^{2/3} M_\star^{1/3}$, $\phi_\star \propto d^{-3} \propto \phi$, and $P\propto \phi^2$.  There is, however, an exception to this trend: when $p_{orb}\approx p_{rot}$, $P$ decreases significantly.  This comes from the $\omega$ calculation.

\subsection{Comparison to Irradiation}
\label{ssec:irradiation_comp}

If the planet is old enough that the internal energy of formation is negligible, the total power deposited into the planet by the star is equal to:

\begin{equation}
    P_{tot} = P_{ind} + P_{lum} - P_{sb},
\end{equation}

\noindent where $P_{lum}$ is the energy incident on the planet caused by irradiation from the host star and $P_{sb}$ is the energy irradiated away from Stefann-Boltzmann radiation.  Re-arranging terms, we can write $P_{tot}$ as:

\begin{equation}
P_{tot} = P_{ind} + \frac{4\sigma \pi R_\star^2 T_\star^4}{4 \pi d^2} (1-a) \pi R_p^2 - 4 \pi \sigma R_p^2 T_{surf}^4,
\end{equation}

\noindent which simplifies to:

\begin{equation}
\label{eqn:ptot_rearranged}
P_{tot} = P_{ind} + (4 \pi \sigma R_p^2)\bigg(  \frac{ R_\star^2 T_\star^4 (1-a) }{4 d^2}   - T_{surf}^4\bigg).
\end{equation}

If the system is in equilibrium and $P_{ind}$ is small, we recover the commonly-used expression for the planetary equilibrium temperature:

\begin{equation}
T_{surf} = T_{eq} =  T_\star \sqrt{\frac{R_\star}{D}} \bigg(\frac{1-A }{4 }\bigg)^{1/4},
\end{equation}

\noindent where $A$ is the albedo.

We can then rewrite Equation~\ref{eqn:ptot_rearranged} as:
\begin{equation}
P_{tot} = P_{ind} + (4 \pi \sigma R_p^2)\bigg(  T_{eq}^4   - T_{surf}^4\bigg).
\end{equation}

Then, we see that there will be a net gain in power if:

\begin{equation}
    P_{ind} > (4 \pi \sigma R_p^2)\bigg( T_{surf}^4 - T_{eq}^4 \bigg),
\end{equation}

\noindent and a net loss in the inverse scenario.  

In general, if $P_{ind}$ is constant with temperature (or increases with temperature more slowly than $R_p^2T^4$), this results in a stable system- the planet gaining power would result in a net gain in energy on the planet's surface, resulting in an increase in temperature $T_{surf}$ and radius $R_p$.  Eventually, the two sides of the equation will equilibrate at some $T_{surf}>T_{eq}$.

\begin{figure}
    \centering
    \includegraphics[width=0.5\linewidth]{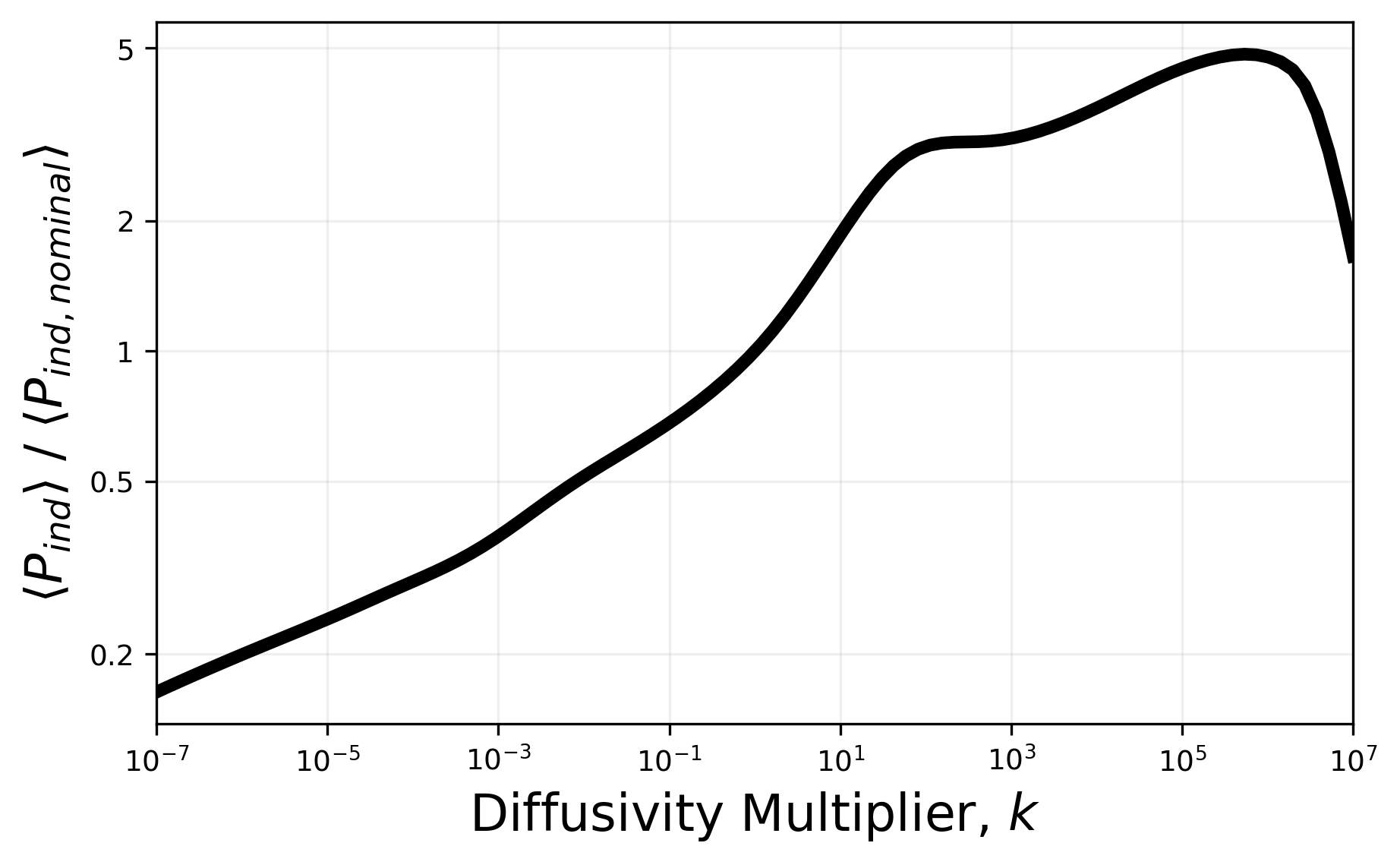}
    \caption{The average power dissipation within a 1\,$M_J$, 3d planet orbiting a sun-like star with a rotation period of 5d and a surface magnetic field of $B_{surf}=1$\,kG as a function of $k$, where $k$ is a multiplicative factor of $\eta$.}
    \label{fig:eta_power}
\end{figure}

However, $P_{ind}$ is not necessarily constant with temperature- as shown in Figure~\ref{fig:ne_nn}, increasing a body's temperature will cause the diffusivity to decrease.  The relationship between diffusivity and total inductive power $P$, as discussed in \cite{Laine2008}, is not a straightforward monotonic relationship- a higher $\eta$ can reduce the generated power-per-volume, but can also increase the area over which the power is dissipated over.  We show, in Figure~\ref{fig:eta_power}, how arbitrarily increasing or decreasing $\eta$ by some multiplicative factor $k$ can influence $P_{ind}$.

Overall, we see that, in the general regime of $\eta$ in which we are operating (where $k=1$), reducing the magnetic diffusivity (which will typically happen when increasing temperature) will $reduce$ the net power dissipation, though the dissipated power decreases by less than an order of magnitude between $k=1$ and $k=10^{-7}$.  When varying $k$ between $10^{-7}$ and $10^7$ (representing an extreme shift in internal conditions across the entire planet), $P_{ind}$ still only changes by a factor of 30.

\begin{figure}
    \centering
    \includegraphics[width=0.5\linewidth]{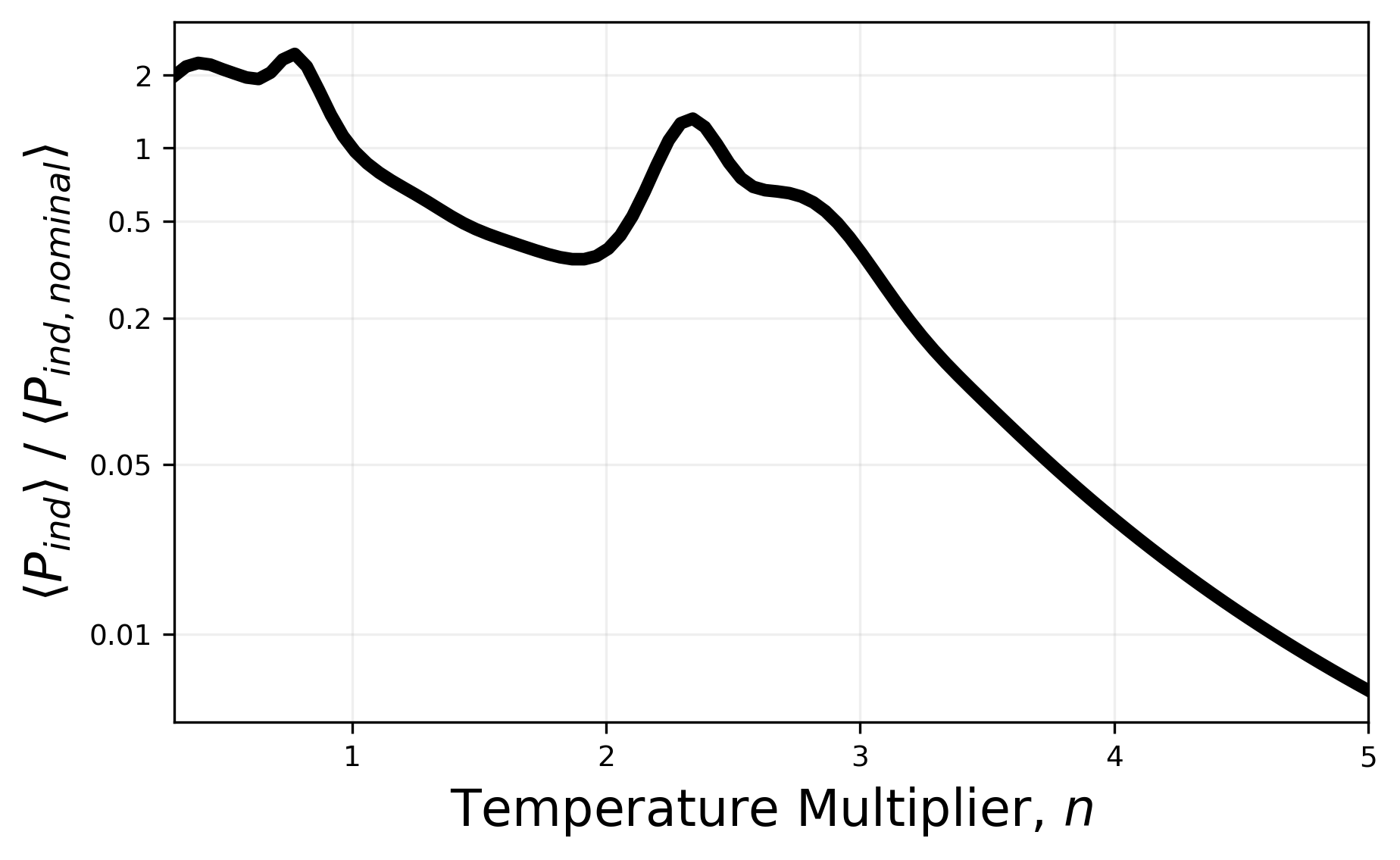}
    \caption{The average power dissipation within a 1\,$M_J$, 3d planet orbiting a sun-like star with a rotation period of 5d and a surface magnetic field of $B_{surf}=1$\,kG as a function of $n$, where $n$ is a multiplicative factor of the temperature.}
    \label{fig:n_power}
\end{figure}

To find the relationship between $T$ and $P_{ind}$, we performed our calculations for the planet with varying temperature profiles, defining a value $n$ such that $T(r) = n T_{GASTLI}(r)$ and varying $n$.  Our results for a 1\,$M_J$ and $p_{orb}=3$d planet are in Figure~\ref{fig:n_power}.  

In general, it appears that, in the starting (that is, $n=1$) case, the planet is in a regime in which increasing the temperature actually results in a slight \textit{decrease} in net power from $P_{ind}$.  Therefore, if we initialize in a scenario in which $P_{ind} > P_{sb} - P_{lum}$, the increase in temperature will both lower $P_{ind}$ and increase $P_{sb}$, resulting in a scenario in which the system equilibriates such that $P_{ind} = P_{sb} - P_{lum}$.  As $P_{sb}$ typically changes rapidly with temperature than $P_{ind}$ around $n=1$, we anticipate that $P_{ind, final} \approx P_{ind, initial}$ and can then calculate the change in planet surface temperature:

\begin{equation}
\label{eqn:tsurf}
    T_{surf} = \bigg(\frac{P_{ind}}{4 \pi \sigma R_p^2} + T_{eq}^4\bigg)^{1/4}.
\end{equation}

To cause an increase in surface temperature of 100\,K on a Jupiter-radius body orbiting a sun-like star at 3$d$ with a zero albedo (and therefore an equilibrium temperature of 1390$K$), we would need $P_{ind} = 4*10^{21}$\,J\,s$^{-1}$.  As shown in Figure~\ref{fig:m_power} (which considers a planet with a similar size and orbital period), this configuration would only be possible if the host star has a surface magnetic moment comparable to a mCP star or a magnetically active giant.  An increase in 10$K$, which would be significantly more difficult to observe (especially given the typical temperature error on stars), would require a $P_{ind} = 4*10^{20}$\,J\,s$^{-1}$, which is still quite high.

We therefore need to search for scenarios in which $P_{ind}$ is expected to be large enough to cause appreciable surface heating.  It is clear from the previous equation that $T_{surf}$ changes the most dramatically when the first term in is large compared to the second term.  That is, we want the ratio:

\begin{equation}
    \frac{P_{ind}}{4 \pi \sigma R_p^2 T_{eq}^4}>1.
\end{equation}

Recall, from previous sections, that we found that $P_{ind}$ scales with $\phi_\star^2$ for most parameters. Therefore, this ratio can be rewritten as omitting the scaling with $R_p$, which is more complex):

\begin{equation}
\label{eqn:scaling}
    \frac{P_{ind}}{4 \pi \sigma  T_{eq}^4} \propto \frac{ B_{surf}^2 R_\star^6 \sin^2(\alpha) }{4 \pi \sigma T_\star^4 \frac{R_\star^2}{16 D^2} (1-A) D^6} \propto \frac{4  B_{surf}^2 R_\star^4  \sin^2(\alpha) }{\pi \sigma T_\star^4  (1-A) D^4} .
\end{equation}

Therefore, the power of inductive heating compared to that from the stellar irradiation is maximized when the surface magnetic field is high, the magnetic field is highly misaligned with the star's rotation axis, the stellar radius is large, the host star's temperature is low, the orbit is close-in, and the albedo is high.  However, these scaling relations assume that the planet has a completely circular orbit- a planet with a higher eccentricity will also have a higher value of $P_{ind}$, which we show in Section~\ref{ssec:ecc}.

\subsection{The Effect of Nonzero Eccentricities}
\label{ssec:ecc}

Planets on eccentric orbits may be able to achieve higher values of $P_{ind}$ than planets on circular orbits, as there are additional changes in the magnetic flux intercepted by the planet as a result of its changing orbital distance.  This should generate additional current inside the planet, resulting in an increase in deposited power.  In this section, we will lay out the framework behind expanding the calculations of \cite{Campbell1983} and \cite{Laine2008} to include the eccentricity of the orbiting body, and will demonstrate the calculations out to first order in eccentricity.

\subsubsection{Re-Evaluating $\phi_\star$ and $\phi_p$}

In a stationary reference frame, the position of a point on the surface of the planet can be described with:

\begin{equation}
\label{eqn:planet_pos}
    (x_p - r \sin(\theta) \cos(\omega_p t + \varphi),x_p - r \sin(\theta) \cos(\omega_p t + \varphi), r \cos(\theta)),
\end{equation}

where $x_p$ and $y_p$ are the $x$ and $y$ positions of the planet in a stationary frame.  If the planet is eccentric:

\begin{equation}
    x_p = D \cos(f), y_p = D\sin(f),
\end{equation}

where $f$ is the true anomaly of the planet and $D$ is the instantaneous distance between the planet and the host star.  To estimate $D$, we utilize Equation 2.83 from \cite{Murray1999} to first order in $e$: 

\begin{equation}
\label{eqn:D_approx}
    \frac{1}{D^3} \approx \frac{1+3 e \cos(M)}{a^3} + O(e^2),
\end{equation}

\noindent where $a$ is the semi-major axis of the planet and $M = 2\pi t/p_{orb}$ is the mean anomaly of the planet.  Let us define $\omega_p = 2\pi/p_{orb}$ as the planet's orbital frequency. Using \cite{Murray1999}, we can also estimate the value of $f$:

\begin{equation}
\label{eqn:f_approx}
    f \approx \omega_p t + 2 e \sin(\omega_p t) + O(e^2).
\end{equation}

To calculate the magnetic field in the frame of the uniformly rotating planet (which we assume rotates synchronously at rate $\omega_p$), we translate this planet position into a uniformly rotating frame by multiplying the planet position in Equation~\ref{eqn:planet_pos} by a rotation matrix around the z-axis with an angle of -$\omega_p t$.  To first order in eccentricity, the planet's position in this new frame (defined with axes the same as in \cite{Campbell1983}, where, the $y$ axis points directly towards the planet and the positive-x direction is toward the right side of the planet when viewed face-on) is:

\begin{equation}
\begin{aligned}
    x' = r \sin(\theta) \sin(\varphi) - 2 a e \sin (\omega_p t); \\
    y' = a - r \sin(\theta) \cos (\varphi) - a e \cos(\omega_p t);\\
    z' = a \cos (\theta); \\
    r'= \sqrt{x'^2 + y'^2 + z'^2} = \sqrt{r^2 + a^2 - 2 r a \sin(\theta) \cos(\varphi)} \\-\frac{a e \bigg(\cos(\omega_p t)\big( a - r \sin\theta) \cos(\varphi)\big) \bigg) + 2 r \sin(\theta)\sin(\varphi) \sin(\omega_p t) }{\sqrt{r^2 + a^2 - 2 r a \sin(\theta) \cos(\varphi)}}.\\
\end{aligned}
\end{equation}

These terms evaluate to the same equations as Equation~\ref{eqn:prime_values} when the planet has a circular orbit ($e=0$, $a=D$).  We then follow the same procedure as performed in Section~\ref{sssec:mag_field} to calculate $\phi_\star$.  For the purposes of simplification, we note that, as described in Section~\ref{ssec:rad_profile}, the $l=2$ terms provide only a minor ($<10\%$) contribution to the total power dissipation compared to the $l=1$ terms, and therefore we will omit them from our calculations in the remainder of this section.  We can therefore write $\phi_\star$ (removing terms of order $e^2$ and higher) as:

\begin{equation}
\label{eqn:phistar_ecc}
\begin{aligned}
    \phi_\star =-\frac{ \mu_0m r^2\cos(\alpha)}{8\pi a^3} P_1^0 \Bigg( 3 e \cos(\omega_pt)  \Bigg) -\frac{ \mu_0m r^2 \sin(\alpha)}{8\pi a^3}P_1^1
    \Bigg( \cos(\omega t) \sin(\varphi)+  2 \sin(\omega t)\cos(\varphi) \\
    + 6 e \cos (\varphi) \sin\big((\omega-\omega_p)t\big) 
    + e \sin(\varphi) \bigg(\frac{9}{2}\cos\big((\omega-\omega_p) t\big) - \frac{3}{2}\cos\big((\omega+\omega_p) t\big) \bigg)
    \Bigg),
\end{aligned}
\end{equation}

\noindent which highlights the new frequency dependence of $\phi_{\star}$.

The additional terms in $\phi_{\star}$ necessitate a new functional form for $\phi_p$.  We now define $\phi_p$:
\begin{equation}
\begin{aligned}
    \phi_p =
     \frac{\mu_0}{r}  P_1^0 \bigg(\epsilon_1 \cos(\omega_p)  + \epsilon_2 \sin(\omega_p) \bigg) \\
    + \frac{\mu_0}{r}  P_1^1\bigg(\alpha_1\sin(\omega t) + \alpha_2 \cos(\omega t)\bigg) \cos (\varphi) 
    + \frac{\mu_0}{r}  P_1^1\bigg(\alpha_3\sin(\omega t) + \alpha_4 \cos(\omega t)\bigg) \sin (\varphi) \\
    + \frac{\mu_0}{r}  P_1^1\bigg(\xi_1\cos(\omega t - \omega_p t) + \xi_2 \cos(\omega t + \omega_p t)\bigg)  \cos (\varphi)
    + \frac{\mu_0}{r}  P_1^1\bigg(\xi_3\sin(\omega t - \omega_p t) + \xi_4 \sin(\omega t + \omega_p t)\bigg)  \cos (\varphi)\\
    + \frac{\mu_0}{r}  P_1^1\bigg(\xi_5\cos(\omega t - \omega_p t) + \xi_6 \cos(\omega t + \omega_p t)\bigg)  \sin (\varphi)
    + \frac{\mu_0}{r}  P_1^1\bigg(\xi_7\sin(\omega t - \omega_p t) + \xi_8 \sin(\omega t + \omega_p t)\bigg)  \sin (\varphi)\\
\end{aligned}
\end{equation}

\subsubsection{Re-Evaluating $\phi_{inner}$}
The new time-dependent terms in $\phi_\star$ and $\phi_p$ indicate that we need to develop a new form for $\phi_{inner}$, that captures a dependence on $\omega_p$.  We propose the following form, which includes terms that vary as $\omega$, $\omega_p$, and both $\omega$ and $\omega_p$:

\begin{equation}
\begin{aligned}
    \phi_{inner} (r, t) = \mu_0 \bigg(\sum_{l, m} C^{m}_{1,l} G_{1,l}(r) Y_l^m(\theta, \varphi)\bigg)e^{i\omega t}  +  \mu_0 \bigg(\sum_{l, m} C^{m}_{1,l} G_{2,l}(r) Y_l^m(\theta, \varphi)\bigg)e^{i\omega_p t}\\ +  \mu_0 \bigg(\sum_{l, m} C^{m}_{3,l} G_{3,l}(r) Y_l^m(\theta, \varphi)\bigg)e^{i(\omega+\omega_p) t}+  \mu_0 \bigg(\sum_{l, m} C^{m}_{4,l} G_{4,l}(r) Y_l^m(\theta, \varphi)\bigg)e^{i(\omega-\omega_p) t} ,
\end{aligned}
\end{equation}

\noindent where $C_{i,l}^m$ are constant coefficients, $G_{i,l}(r)$ are arbitrary functions, and $Y_l^m(\theta, \varphi)$ are spherical harmonics.  Similarly to the zero-eccentricity case, we can calculate $G_{i,l}(r)$ and $C_{i,l}^m$ separately.

First, we must consider how we would find the three separate $G_{i,l}(r)$ terms.  For the sake of simplicity, we define $\phi_1$, $\phi_2$, $\phi_3$, and $\phi_4$, where:

\begin{equation}
\begin{aligned}
\phi_1 = \mu_0 \bigg(\sum_{l, m} C^{m}_{1,l} G_{1,l}(r) Y_l^m(\theta, \varphi)\bigg)e^{i\omega t},~~\phi_2 = \mu_0 \bigg(\sum_{l, m} C^{m}_{2,l} G_{2,l}(r) Y_l^m(\theta, \varphi)\bigg)e^{i\omega_p t} ,~~\\ \phi_3 = \mu_0 \bigg(\sum_{l, m} C^{m}_{3,l} G_{3,l}(r) Y_l^m(\theta, \varphi)\bigg)e^{i(\omega +\omega_p) t}, ~~  \phi_4 = \mu_0 \bigg(\sum_{l, m} C^{m}_{4,l} G_{4,l}(r) Y_l^m(\theta, \varphi)\bigg)e^{i(\omega-\omega_p) t},
\end{aligned}
\end{equation}

\noindent such that $\phi_{inner} = \phi_1 + \phi_2$ + $\phi_3$ + $\phi_4$.

Inserting our value for $\phi_{inner}$ into Equation~\ref{eqn:b_phi}, we find:

\begin{equation}
\label{eqn:split_phi}
    \nabla \times B = -\frac{i \omega }{\eta} \nabla \times \phi_1   -\frac{i \omega_p}{\eta} \nabla \times \phi_2    -\frac{i (\omega +  \omega_p)}{\eta} \nabla \times \phi_3   -\frac{i( \omega - \omega_p)}{\eta} \nabla \times \phi_4.
\end{equation}

We find:

\begin{equation}
    \frac{d^2 G_{1,l}}{d r^2}(r) - \Bigg(\frac{l(l+1)}{r^2} + \frac{i\omega}{\eta}\Bigg) G_{1,l}(r) = 0,
\end{equation}

\begin{equation}
    \frac{d^2 G_{2,l}}{d r^2}(r) - \Bigg(\frac{l(l+1)}{r^2} + \frac{i\omega_p}{\eta}\Bigg) G_{2,l}(r) = 0,
\end{equation}

\begin{equation}
    \frac{d^2 G_{3,l}}{d r^2}(r) - \Bigg(\frac{l(l+1)}{r^2} + \frac{i (\omega + \omega_p)}{\eta}\Bigg) G_{3,l}(r) = 0,
\end{equation}

\noindent and

\begin{equation}
    \frac{d^2 G_{4,l}}{d r^2}(r) - \Bigg(\frac{l(l+1)}{r^2} + \frac{i( \omega - \omega_p)}{\eta}\Bigg) G_{4,l}(r) = 0,
\end{equation}

\noindent assuming the terms with different $e^{i\omega t}$ and $e^{i\omega_p t}$ dependencies are independent.

We can then find the different values of $G_{i,l}$ using the same methods as we did for the circular-orbit case, except with different values of $\omega$.  

\subsubsection{Calculating the Power}

We now must find the $C_{i,l}^m$ coefficients.  A detailed description of how we calculated these coefficients is included in Appendix~\ref{app:clm_ecc}.

After solving for all of the $\mu$ and $\nu$ values, we can finally write out the full form of $\phi_{1}$, $\phi_{2}$, $\phi_{3}$, and $\phi_4$:

\begin{equation}
\begin{aligned}
    \phi_{inner} (r, t) = \mu_0\Bigg( \bigg(-(\mu_{1,1}^1 + i \nu_{1,1}^1) G_{1,1}(r) \sqrt{\frac{3}{8\pi}} \sin(\theta) e^{i\varphi} + (\mu_{1,1}^{-1} + i \nu_{1,1}^{-1})  G_{1,1}(r) \sqrt{\frac{3}{8\pi}} \sin(\theta) e^{-i\varphi}\bigg)e^{i\omega t}  \\
    +   \bigg( (\mu_{2,1}^0 + i \nu_{2,1}^0) G_{2,1}(r)\sqrt{\frac{3}{4\pi}} \cos(\theta)\bigg)e^{i\omega_p t} \\
    +  \bigg(-(\mu_{3,1}^1 + i \nu_{3,1}^1) G_{3,1}(r)\sqrt{\frac{3}{8\pi}} \sin(\theta) e^{i\varphi}+  (\mu_{3,1}^{-1} + i \nu_{3,1}^{-1}) G_{3,1}(r) \sqrt{\frac{3}{8\pi}} \sin(\theta) e^{-i\varphi}\bigg)e^{i(\omega+\omega_p)t}  \\
    +  \bigg(-(\mu_{4,1}^1 + i \nu_{4,1}^1) G_{4,1}(r)\sqrt{\frac{3}{8\pi}} \sin(\theta) e^{i\varphi}+  (\mu_{4,1}^{-1} + i \nu_{4,1}^{-1}) G_{4,1}(r) \sqrt{\frac{3}{8\pi}} \sin(\theta) e^{-i\varphi}\bigg)e^{i(\omega-\omega_p) t}  
    \Bigg).
\end{aligned}
\end{equation}

We can then estimate $P_{ind}$ by inserting this equation into the integral from Equation~\ref{eqn:pind_integral}.  Recall that:

\begin{equation}
    P_{ind} = \iiint \frac{\eta (r) }{\mu_0}\,\big(\mathrm{Re}(\nabla \times B)\big)^2 r^2 \sin(\theta)  dr d\theta d\varphi,
\end{equation}

\noindent and, using Equation~\ref{eqn:split_phi}:

\begin{equation}
\begin{aligned}
    P_{ind} = \iiint \frac{1}{\mu_0 \eta (r)} \,\Bigg(\mathrm{Re}\bigg(-\frac{i \omega }{\eta} \nabla \times \phi_1   -\frac{i \omega_p}{\eta} \nabla \times \phi_2    -\frac{i (\omega +  \omega_p)}{\eta} \nabla \times \phi_3 - \frac{i (\omega -  \omega_p)}{\eta} \nabla \times \phi_4 
\bigg)\Bigg)^2r^2 \sin(\theta)  dr d\theta d\varphi,
\end{aligned}
\end{equation}

\noindent which we can write as:

\begin{equation}
\begin{aligned}
    P_{ind} = \iiint \frac{\sin(\theta) }{\mu_0\eta(r)}\,\Bigg[ \sum_{i=1}^4   \omega_i \Bigg(\bigg(\frac{1}{\sin(\theta)} \frac{\partial \mathrm{Im}(\phi_i)}{\partial \varphi} \bigg) \mathbf{\hat{\theta}} - \bigg(\frac{\partial \mathrm{Im}(\phi_i)}{\partial \theta}\bigg) \hat{\varphi}\Bigg) 
    \Bigg]^2 dr d\theta d\varphi,
\end{aligned}
\end{equation}

\noindent or:

\begin{equation}
\label{eqn:pind_splitphi}
\begin{aligned}
    P_{ind} = \iiint \frac{\mu_0\sin(\theta) }{\eta(r)}\,\Bigg(\bigg(\sum_{i=1}^4\frac{\omega_i}{\sin(\theta)} \frac{\partial \mathrm{Im}(\phi_i)}{\partial \varphi}  \bigg)^2
    +\bigg(\sum_{i=1}^4  \omega_i \frac{ \partial \mathrm{Im}(\phi_i)}{\partial \theta} \bigg)^2 \Bigg) dr d\theta d\varphi,
\end{aligned}
\end{equation}

\noindent where $\omega_1 = \omega$, $\omega_2 = \omega_p$, $\omega_3 = \omega + \omega_p$, and $\omega_4 = \omega-\omega_p$.  We therefore need to find Im($\phi_{1}$), Im($\phi_2$), Im($\phi_3$), and Im($\phi_4$).  We find, that for $i=1, 3, 4$:

\begin{equation}
\begin{aligned}
\mathrm{Im}(\phi_{i})=
\sqrt{\frac{3}{8\pi}}\sin(\theta)\Bigg(
\cos(\omega_i t) \cos(\varphi) \big( \mathrm{Re}(G_{i,1})(-\nu_{i,1}^1+\nu_{i,1}^{-1})  + \mathrm{Im}(G_{i,1})(-\mu_{i,1}^1+\mu_{i,1}^{-1}) \big)\\
+\cos(\omega_i t) \sin(\varphi) \big( \mathrm{Re}(G_{i,1})(-\mu_{i,1}^1-\mu_{i,1}^{-1})  + \mathrm{Im}(G_{i,1})(\nu_{i,1}^1+\nu_{i,1}^{-1}) \big)\\
+\sin(\omega_i t) \cos(\varphi) \big( \mathrm{Re}(G_{i,1})(-\mu_{i,1}^1+\mu_{i,1}^{-1})  + \mathrm{Im}(G_{i,1})(\nu_{i,1}^1-\nu_{i,1}^{-1}) \big)\\
+\sin(\omega_i t) \sin(\varphi) \big( \mathrm{Re}(G_{i,1})(\nu_{i,1}^1+\nu_{i,1}^{-1})  + \mathrm{Im}(G_{i,1})(\mu_{i,1}^1+\mu_{i,1}^{-1}) \big)\Bigg),
\end{aligned}
\end{equation}

\noindent and, for $i=2$:

\begin{equation}
\begin{aligned}
\mathrm{Im}(\phi_{2})=
\sqrt{\frac{3}{4\pi}}\cos(\theta)\Bigg(\cos(\omega_p t) \big( \mathrm{Re}(G_{2,1})\nu_{2,1}^0 + \mathrm{Im}(G_{2,1}) \mu_{2,1}^0   \big)+
\sin(\omega_p t) \big( \mathrm{Re}(G_{2,1})\mu_{2,1}^0 - \mathrm{Im}(G_{2,1}) \nu_{2,1}^0   \big)\Bigg).
\end{aligned}
\end{equation}

Inserting these value into Equation~\ref{eqn:split_phi}, we can find $P_{ind}$.  The final equation is very lengthy, so we do not reproduce it here, especially as many of the terms already have to be estimated numerically.  We provide a code that implements these calculations online\footnote{https://github.com/madisontbrady/ohmic\_heating}.

\subsubsection{The Influence of Eccentricity on $P_{ind}$}

\begin{figure}
    \centering
    \includegraphics[width=1.0\linewidth]{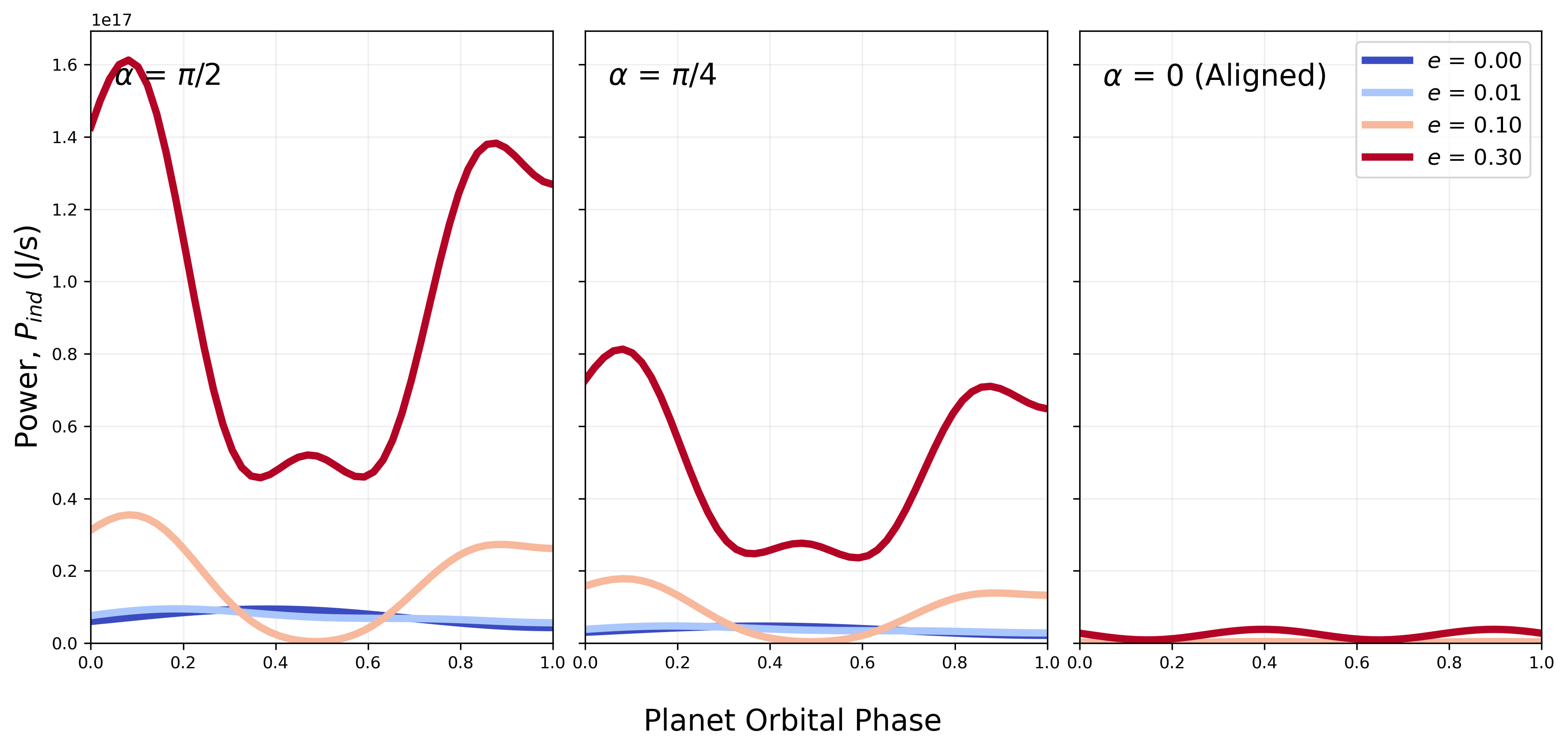}
    \caption{$P_{ind}$ as a function of orbital phase for a 1\,$M_J$ planet on a 3d orbit around a $B_{surf}=1000$G sun-like star with a 5d rotation period for various different values of $\alpha$ and $e$.  $P_{ind}$ was calculated with an approximation of $D$ out to first order in $e$, and therefore may not be fully accurate, especially in the $e=0.3$ scenario.}
    \label{fig:pind_vs_phase_ecc_alpha}
\end{figure}

We can now numerically estimate how the eccentricity influences $P_{ind}$.  In Figure~\ref{fig:pind_vs_phase_ecc_alpha}, we show the $e$ influences the $P_{ind}$ of a 3d planet both as a function of time and $\alpha$.  A planet with a zero eccentricity has a $P_{ind}$ that varies like a sinusoidal curve with time, and increasing $e$ causes this curve to deviate more and more from this configuration.  In addition, a nonzero eccentricity enables $P_{ind}$ to be nonzero even when the planet is aligned with the rotation axis of its host star ($\alpha=0$)- this makes sense, as a nonzero eccentricity results in the planet experiencing a time-dependent magnetic field as a result of the changing orbital distance.

\begin{figure}
    \centering
    \includegraphics[width=1.0\linewidth]{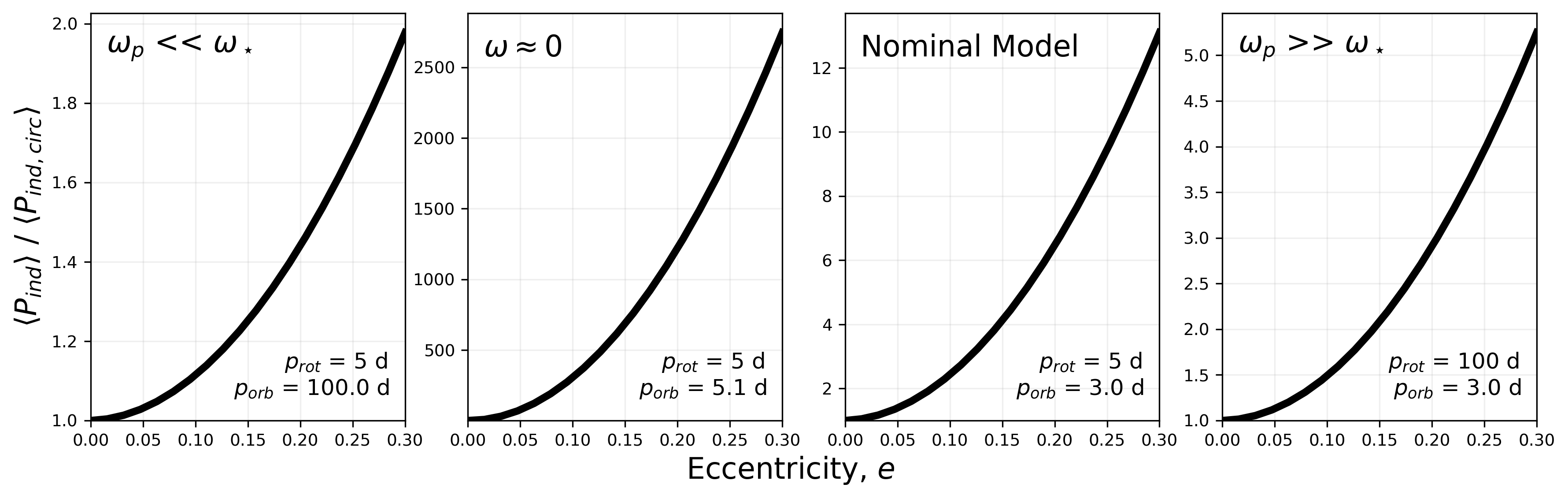}
    \caption{The average change in $P_{ind}$ as a function of eccentricity for a 1\,$M_J$ planet in various different orbital configurations around a $B_{surf}=1000$G sun-like star.}
    \label{fig:pind_ecc}
\end{figure}

Figure~\ref{fig:pind_ecc} shows how the average $P_{ind}$ on the planet over one orbital period is influenced by the planet's eccentricity for various different values of $p_{orb}$.  We note that there is no simple scaling relationship between $P_{ind}$ and the eccentricity, but instead that this relationship changes in a complex manner as a function of $\omega/\omega_p$, which may necessitate calculations on a case-by-case basis. It is also important to re-iterate that our calculations in this section only consider terms out to first order in eccentricity.  This means that our estimations of $P_{ind}$ are unlikely to be accurate for large values of $e$.  However, that being said, we can see that, in most cases (besides the unique situation where $\omega\approx0$, in which case the circular-orbit $P_{ind}\approx 0$) low values of eccentricity ($e<0.2$) do not seem to have a large influence on $P_{ind}$, causing increases by a factor of two or less.  Therefore, for most old short-period planets, which are expected to have low values of $e$ due to tidal effects, the planetary eccentricity is unlikely to have a significant effect on $P_{ind}$.

We note that our above results assume that $|\omega_p|$, $|\omega|$, $|\omega-\omega_p|$, and $|\omega+\omega_p|$ are all unique values.  If, for example, $p_{orb} = 3/2 p_{rot}$, $|\omega-\omega_p|$ = $\omega$, the terms dependent on $\omega-\omega_p$ and $\omega$ in Equation~\ref{eqn:phistar_ecc} are not independent from one another and therefore must be combined coherently.  Moreover, the total power calculation in Equation~\ref{eqn:pind_splitphi} assumes that the $\omega_i$ terms are independent, which is incorrect in this particular scenario, where $|\omega_1| = |\omega_4|$.  As most planets do not appear to have orbits synchronized in this manner with their host stars, we omit the full numerical calculation of this scenario in this paper.  However, it is straightforward to modify our math to combine the $\omega_1$ and $\omega_4$ terms in our $C_l^m$ coefficient and power calculations to account for a planet on a commensurate orbit.

It is straightforward, but time-consuming, to expand our methods in this section out to additional orders of $e$.  To do so, we would extend our $D$ approximation in Equation~\ref{eqn:D_approx} and $f$ approximation in Equation~\ref{eqn:f_approx} out to additional orders of $e$.  For example, to second-order \citep[following][]{Murray1999}:

\begin{equation}
    D \approx a \big( 1 - e \cos(\omega_p t) + e^2 - e^2 \cos(2 \omega_p t) \big),
\end{equation}

\noindent and:

\begin{equation}
\label{eqn:approx_wframe_cos}
\begin{aligned}
    f \approx w_p t + 2 e \sin(\omega_p t) + 5 e^2 \cos(\omega_p t) \sin(\omega_p t)
\end{aligned}
\end{equation}

Inserting these approximations into $\phi_\star$ would result in the creation of additional periodic terms that vary as, e.g., $\sin(\omega t \pm 2 \omega_p t)$, which would necessitate the addition of new terms in $\phi_p$ and $\phi_{inner}$ that also vary like $(\omega \pm 2 \omega_p) t$.  We would then have to calculate the $G$ functions and $C$ coefficients for these additional terms in $\phi_{inner}$ in order to calculate $P_{ind}$.  As the majority of short-period planets have low eccentricities, we will leave the detailed analytical expansion out to higher orders in eccentricity to future works.

\subsection{The Effect of Nonzero Obliquities}
\label{ssec:obl}

As discussed in \cite{Chang2012}, a planet with a nonzero obliquity will also experience a changing magnetic field.  We can model this by performing the same calculations as we did in Section~\ref{sec:methods}, except with a replacement of $\hat{m}$ in Equation~\ref{eqn:mhat_defn}:

\begin{equation}
\begin{aligned}
    \mathbf{\hat{m}} = \bigg(\sin(\alpha) \big(\cos(\omega_\star t) \cos(\omega_p t) + \cos(\lambda) \sin(\omega_\star t)   \sin(\omega_p t)\big) + \sin(\lambda) \cos(\alpha) \sin(\omega_p t)\bigg) \mathbf{i'}\\
    + \bigg(\sin(\alpha) \big(-\cos(\omega_\star t) \sin(\omega_p t) + \cos(\lambda) \sin(\omega_\star t)   \cos(\omega_p t)\big) + \sin(\lambda) \cos(\alpha) \cos(\omega_p t) \bigg)\mathbf{j'} \\
    +\bigg(- \sin(\lambda) \sin(\alpha) \sin(\omega_\star t) + \cos(\lambda) \cos(\alpha) \bigg) \mathbf{k'},
\end{aligned}
\end{equation}

\noindent where $\lambda$ is the orbital obliquity and $\omega_\star = 2\pi/p_{rot}$.  While \cite{Chang2012} performs the full calculation in the case of free $\alpha$ and $\lambda$, we are interested in determining the upper limit on $P_{ind}$, so we will merely highlight one limiting case of $\alpha$ in these calculations.  Consider the situation in which $\alpha=0$.  In this case, we find:
\begin{equation}
\begin{aligned}
    \mathbf{\hat{m}} = \bigg( \sin(\lambda) \sin(\omega_p t)\bigg) \mathbf{i'}
    + \bigg( \sin(\lambda) \cos(\omega_p t) \bigg)\mathbf{j'} 
    +\bigg(\cos(\lambda)  \bigg) \mathbf{k'},
\end{aligned}
\end{equation}

From this, we find the non-constant, non-spatially independent portion of $\phi_\star$ (up to l=1):

\begin{equation}
\label{eqn:phistar_obl_alphazero}
    \phi_\star = -\frac{m  \sin(\lambda) r^2}{8 \pi D^3} P_1^1\bigg(2 \cos(\omega_p t) \cos(\varphi)  + \sin(\omega_p t) \sin(\varphi)  \bigg).
\end{equation}

As this function has a very similar form to $\phi_\star$ in the zero-eccentricity case (in Equation~\ref{eqn:phi_star}), we can use the same formulation for $\phi_p$ and $\phi_{inner}$ as in Section~\ref{sec:methods}, with the only difference being that we replace $\omega$ with $\omega_p$.  The process for finding $G_l(r)$ for $\phi_{inner}$ is identical (except for the substitution of $\omega = \omega_p$), so we will only summarize the calculation of the $C_l^m$ coefficients in Appendix~\ref{app:clm_lam}.  As $\phi_{inner}$ has the same form as in the nominal case described in Section~\ref{sec:methods}, we can calculate $P_{ind}$ using the same techniques as in Section~\ref{sec:methods}, but with different values of $C_l^m$ (and, therefore, different exact values of $A$) and using $\omega_p$ in place of $\omega$.

\begin{figure}
    \centering
    \includegraphics[width=0.5\linewidth]{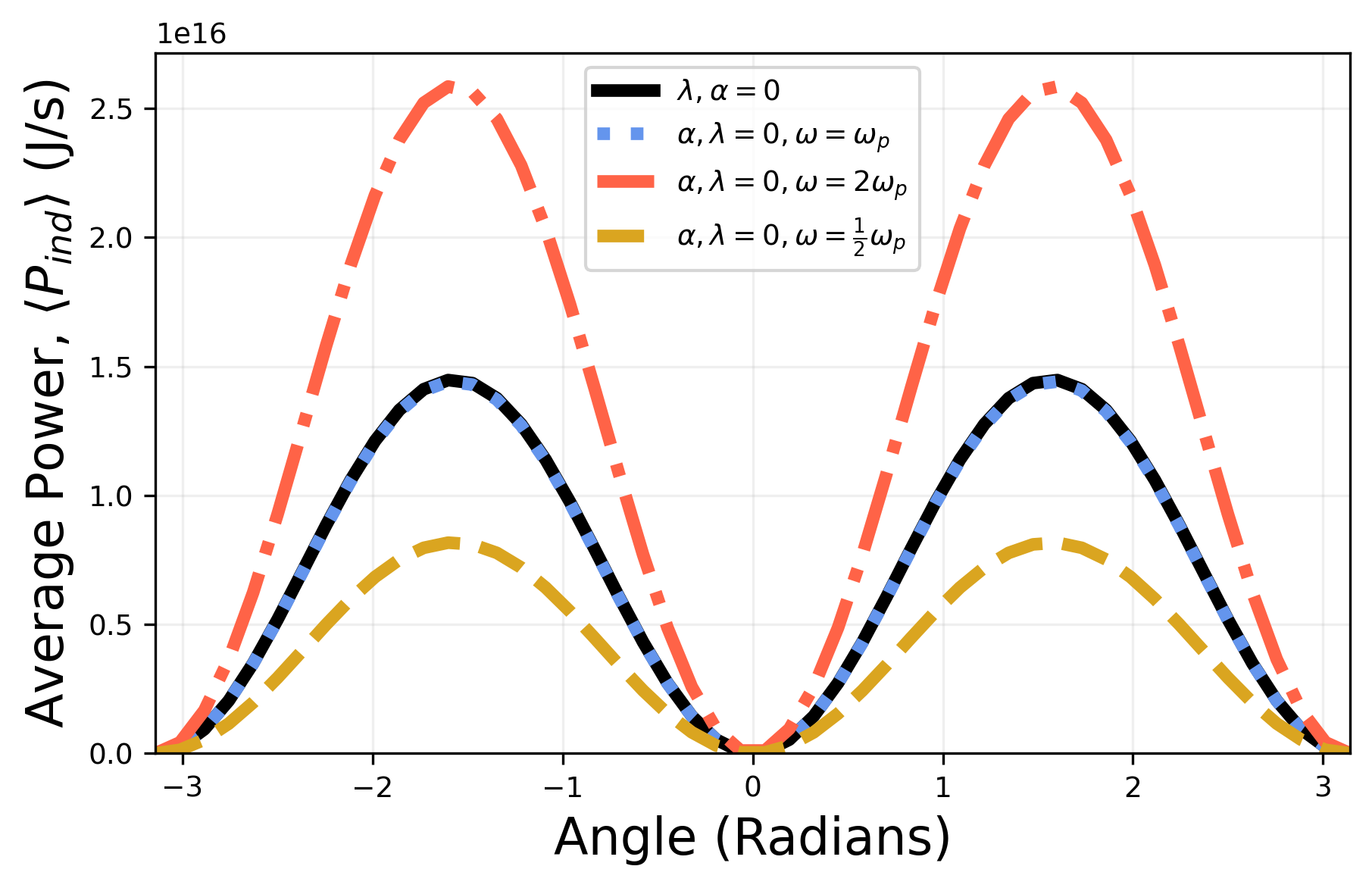}
    \caption{The average power $\langle P_{ind}\rangle$ as a function of $\lambda$ and $\alpha$.  The solid line indicates the power as a function of $\lambda$ when $\alpha$=0, while the dotted and dashed lines indicate the power as a function of $\alpha$ when $\lambda=0$ for various values of $\omega = \omega_\star - \omega_p$.}
    \label{fig:angle_power}
\end{figure}

In Figure~\ref{fig:angle_power}, we compare our results for $\langle P_{ind} \rangle$ for varying values of $\lambda$ in the $\alpha=0$ scenario compared to our results for the same value while varying $\alpha$ in the $\lambda=0$ scenario.   We found that allowing $\omega=\omega_p$ in the zero-obliquity scenario resulted in $\langle P_{ind} \rangle$ values very similar to those in the zero-$\alpha$ scenario, with the only differences appearing to be the phase of the variations in $P_{ind}$.  We found that the average power scales as $\sin(\lambda)^2$, similarly to how it scales with $\alpha$ in the zero-obliquity case.  This makes sense, as $\phi_\star$ scales like $\sin(\lambda)$ in Equation~\ref{eqn:phistar_obl_alphazero}.  Therefore, it appears that we can perform a reasonable estimate of $\langle P_{ind} \rangle$ for a planet on a misaligned orbit by utilizing our prescription from Section~\ref{sec:methods}, but replacing $\omega$ with $\omega_p$ and $\alpha$ with $\lambda$.  However, this estimate only works in the scenario that $\alpha = 0$ (that is, the stellar rotation axis is aligned with its magnetic dipole).

Finding $P_{ind}$ in the case of both a nonzero $\alpha$ and $\lambda$ is considerably more complex.  However, we note that \cite{Chang2012} calculated $P_{ind}$ for several different discrete values of $\alpha$ and $\lambda$, and it appears that varying $\lambda$ does not appear to change $P_{ind}$ by more than a factor of two when $\alpha = \pi/2$.  Therefore, the upper limits that we calculate on $P_{ind}$ with our model in the $\alpha=\pi/2$ limit are likely still a reasonable estimate of the upper limit on $P_{ind}$, even though we are not taking obliquity into account.  We leave a combined analysis involving estimating $P_{ind}$ for nonzero $\alpha$, eccentricity, and $\lambda$ to future works.

\section{Comparison To Observed Systems}
\label{sec:comp_to_exo}

Now that we understand how to calculate $P_{ind}$ and how the various input parameters influence $P_{ind}$, we can now estimate $P_{ind}$ for the observed transiting exoplanet population and determine whether or not there are any planets for which $P_{ind}$ is high enough to meaningfully influence their observable characteristics.  

Unfortunately, our model requires us to know the equilibrium temperatures, atmospheric metallicities, CMFs, C/O ratios, internal temperatures, and masses of exoplanets in order to use \texttt{GASTLI} to estimate their temperature-pressure profiles.   While the equilibrium temperatures are simple to calculate for most transiting planets, the remaining parameters are often completely unconstrained.  Therefore, in this section, when using \texttt{GASTLI}, we will used fixed values for the metallicity (solar), CMF (0.01), C/O ratio (solar), internal temperatures (100\,K) and mass (1\,$M_J$), and simply rescale the profiles to match the observed bulk radius of each planet.  While this is certainly not going to completely accurately reproduce the true temperature-pressure profiles of each planet, a comprehensive \texttt{GASTLI} analysis of all known exoplanets is beyond the scope of this work.  In addition, we do not anticipate $P_{ind}$ to be a strong function of $\eta$ (which is the only factor in our analysis influenced by the temperature-pressure profile), as shown in Figure~\ref{fig:eta_power}- changing $\eta$ by several orders of magnitude typically changes $P_{ind}$ by a factor of a few.

We note that our nominal $\eta$ model (which assumes that the primary source of conductivity in these planets is caused by Saha ionization) also may not be an accurate representation of the conductivity in the upper atmospheres of cold planets (especially those where $p_{orb}>30$d or those orbiting low-luminosity stars, where even alkali metals rarely ionize and other electron sources may dominate).  Given the weak relationship between $P_{ind}$ and $\eta$, leave more accurate models of cold planet conductivity to future works.  Given the sharp decrease in $P_{ind}$ with planetary orbital period (see Figure~\ref{fig:porb_power}), we do not anticipate that long-period planets will have observable heating signatures even if they have additional non-Saha sources of atmospheric electrons.  However, the situation is less certain when considering low-luminosity planet hosts (such as pulsars and white dwarfs), where the planet can have a low surface temperature while still being very close-in.

We downloaded the list of confirmed exoplanets from the NASA Exoplanet Archive (accessed August 26, 2026), selected the default parameter sets for each planet, and used our models to estimate $P_{ind}$ for every body for which $R_p>4~{R_\oplus}$ (selecting for planets that are likely to have thick, gas-rich atmospheres) and for which $p_{orb}$, $R_\star$, $M_\star$, and $T_\star$ are known.   We eliminated targets for which $R_p>10R_J$, assuming that such targets are unlikely to be planets.  We also eliminated targets for which $p_{orb}>1000$d, as such planets likely have orbital periods far too long for $P_{ind}$ to be large.  We note that, in most cases, $B_{surf}$, $p_{rot}$, and $\alpha$ are not known.  Therefore, to facilitate comparison between the different observed systems, we fixed these values at numbers that we believe will give us a reasonable upper limit on $P_{ind}$, given the system's properties, with $\alpha$=$\pi/2$ (which would maximize $P_{ind}$) and $p_{rot}=0.1$d, which is around the shortest rotation period observed in the sample of 713 M dwarfs in \citet{Popinchalk2021}.  We note that this rotation period is likely lower than is realistic for solar-type stars (as it results in rotation speeds near or at the stellar break-up velocity), and therefore the estimated $P_{ind}$ is an even more extreme upper limit for early-type stars.  When considering the magnetic field strength $B_{surf}$, we separate the planet hosts into five different categories:

\begin{itemize}
    \item Giant stars.  Inspired by \cite{Stassun2018}, we defined giant stars as those with log$g$\,$<$\,3.5.  In this case, we assumed magnetic fields are giant-like, with an upper limit on $B_{surf}$ of 100\,G as found in \cite{Auriere2015}.  We note that \cite{Stassun2018} also defines a category of sub-giants (with 3.5\,$<$\,log$g$\,$<$\,4.1), but we will group those stars in with main sequence stars (which typically have higher magnetic fields) in order to estimate upper limits on $P_{ind}$.
    \item M dwarfs.  Of the dwarf stars (with log$g$\,$\geq$\,3.5), we selected the stars with $T_{eff}\,<$\,4000\,K as M dwarfs.  These have higher surface magnetic fields- in this case, we set a limit of 10000\,G, which is in-line with the highest magnetic fields observed on M dwarfs in \cite{Reiners2022}.  
    \item Other main sequence stars.  For stars with log$g$\,$\geq$\,3.5 and $T_{eff}\,\geq$\,4000\,K, we set a limit on $B_{surf}$ of 1000\,G, in line with what has been observed on young sun-like stars in \cite{Kochukhov2020}.
    \item Neutron stars.  For neutron star hosts, we used the list of objects described in \cite{Laycock2025}, many of which are not included in the NASA archive and therefore require individual study.  Interestingly, all neutron stars with observed planets (or sub-stellar companions) are pulsars with extremely well-characterized rotation periods, enabling us to directly use the measured $p_{rot}$ values instead of assuming a constant value for all planet hosts.
    \item White dwarfs.  We separated out white dwarf hosts by identifying stars with $R_\star\,<\,0.05$\,$R_\odot$.  Given the small number of confirmed exoplanets orbiting white dwarfs, we were able to perform a literature search on each target in the list and identified all that are known white dwarf planets.  
\end{itemize}

We will discuss results for the first three categories in this section, and stellar remnants in the following subsections.

We calculated the average $P_{ind}$ over one orbital period for each planet given its radius, orbital period, and host star properties.  We calculated $P_{ind}$ for both the $\eta$ model from \cite{Laine2008} and our own alkali-metal prescription included in Section~\ref{ssec:diffusivity}, and quoted the higher of the two values to provide the maximum upper limit.  Given the likely inaccuracy of our low-order eccentricity approximation for $P_{ind}$, we assumed a zero eccentricity for all planets.  We note that small-but-nonzero ($<0.1$) eccentricities are unlikely to have a dramatic effect on our results (see Figure~\ref{fig:pind_ecc}), so this estimate should be reasonable for all systems except for those that have high eccentricities, which may have higher values of $P_{ind}$.

\begin{figure}
    \centering
    \includegraphics[width=1\linewidth]{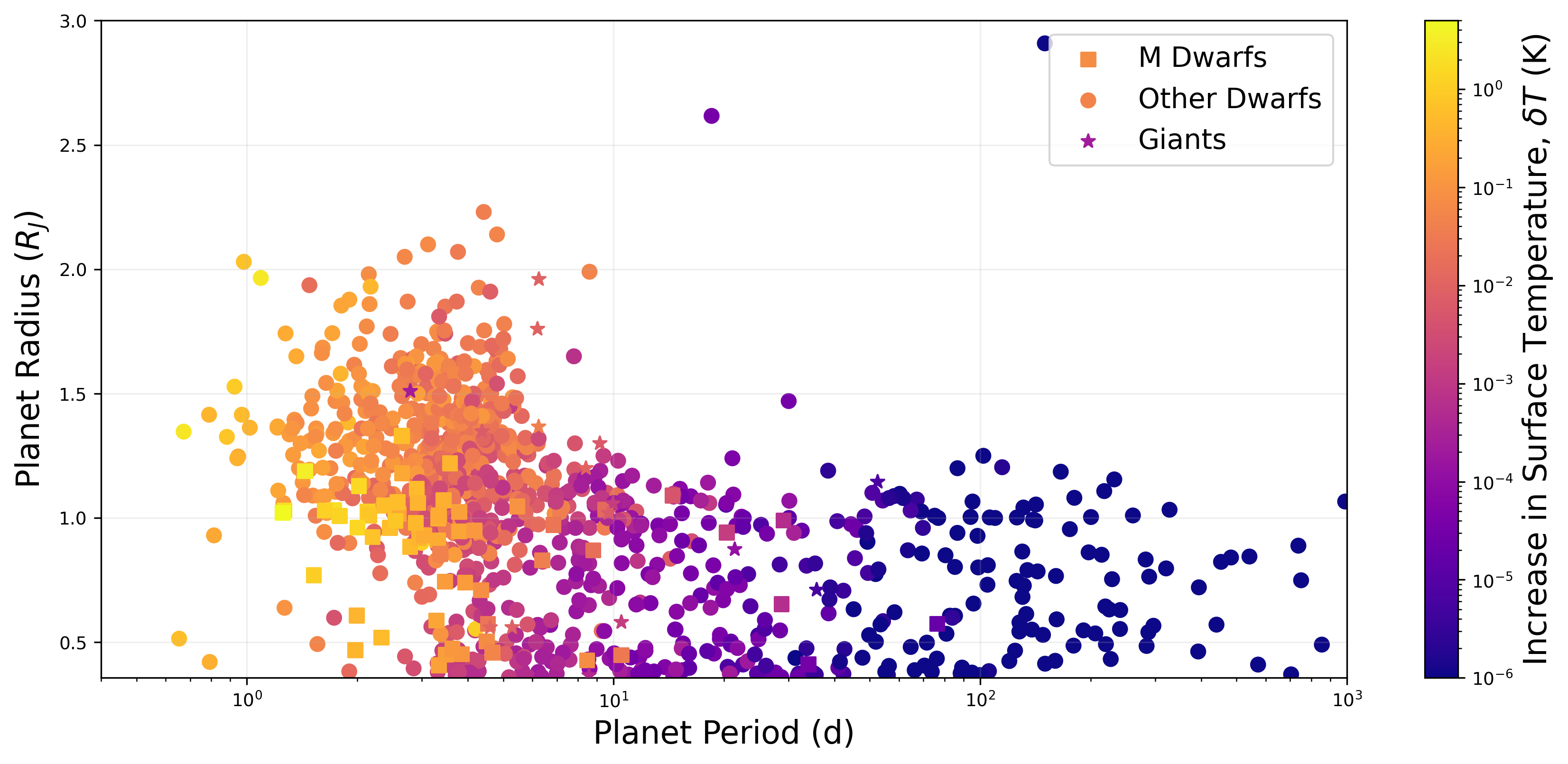}
    \caption{The population of exoplanets with $R_p>4\,R_\oplus$ from the NASA exoplanet archive, colored according to our upper-limit $\delta T$ values.  For each planet, we calculated $\delta T$ using the $P_{ind}$ value for both of our different $\eta$ prescriptions and quoted the higher value.  Planets with M dwarf hosts are shown as squares, planets with giant hosts are stars, and all others are shown as circles.}
    \label{fig:deltaT_rp}
\end{figure}

We were unable to identify any planets with $\delta T$ values higher than 10\,K, which we identify as being around the precision that we would be capable of detecting changes in surface temperature.  We show all planets with $\delta T\,>\,1$K in Table~\ref{tab:exo_pind}, and all of the planets colored according to $\delta T$ in Figure~\ref{fig:deltaT_rp}. We note that these values of $\delta T$ are likely upper limits unless the host star possesses extremely unusual properties (such as having an extremely short rotation period or a superlatively high surface magnetic field) or the planet is has a very high eccentricity.

\begin{table*}
    \centering
    \begin{tabular}{|cccccccccccc|}
    \hline
         Name & $p_{orb} $(d) & $R_p$ ($R_J$) & $R_\star$ ($R_\odot$) & $M_\star$ ($M_\odot$) & $T_\star$ (K) & $T_{eq}$ (K) & $P_{ind}$ (J/s) & $\delta T$ (K) & $\eta$  & Ref. \\
         \hline
TOI-7394 B b & 1.25 & 1.02 & 0.61 & 0.62 & 3717 & 1003 & 9.79E+19 & 6.34 & Laine & \cite{Premnath2026}\\ 
TIC 46432937 b & 1.44 & 1.19 & 0.53 & 0.56 & 3572 & 873 & 4.62E+19 & 3.36 & Laine & \cite{Hartman2024}\\ 
WASP-12 b & 1.09 & 1.97 & 1.69 & 1.32 & 6265 & 2601 & 2.67E+21 & 2.70 & Laine & \cite{Leonardi2024}\\ 
TOI-2109 b & 0.67 & 1.35 & 1.70 & 1.45 & 6540 & 3149 & 2.06E+21 & 2.49 & Laine &\cite{Wong2021}  \\ 
HATS-74 A b & 1.73 & 1.03 & 0.58 & 0.60 & 3777 & 895 & 2.19E+19 & 1.96 & Laine & \cite{Jordan2022}\\ 
TOI-519 b & 1.27 & 1.03 & 0.35 & 0.34 & 3322 & 751 & 1.28E+19 & 1.94 & Laine & \cite{Kagetani2023}\\ 
TOI-5292 A b & 2.02 & 1.13 & 0.61 & 0.60 & 3919 & 910 & 2.19E+19 & 1.56 & Laine & \cite{Dransfield2026}\\ 
TOI-5616 b & 2.00 & 0.96 & 0.65 & 0.67 & 3996 & 939 & 1.65E+19 & 1.47 & Laine & \cite{Kanodia2024}\\ 
TOI-6383 A b & 1.79 & 1.01 & 0.46 & 0.46 & 3444 & 752 & 8.04E+18 & 1.27 & Laine & \cite{Bernabo2024}\\ 
TOI-2337 b & 2.99 & 0.90 & 3.22 & 1.32 & 4780 & 1957 & 1.02E+20 & 1.16 & Laine & \cite{Grunblatt2022}\\ 
TOI-5205 b & 1.63 & 1.03 & 0.39 & 0.39 & 3430 & 737 & 6.96E+18 & 1.12 & Laine & \cite{Kanodia2023} \\ 
TOI-2384 b & 2.14 & 1.02 & 0.61 & 0.64 & 3943 & 890 & 1.20E+19 & 1.11 & Laine &\cite{Bryant2024} \\ 
TOI-4860 b & 1.52 & 0.77 & 0.36 & 0.34 & 3260 & 699 & 3.19E+18 & 1.08 & Laine & \cite{Almenara2024}\\ 
TIC 365102760 b & 4.21 & 0.55 & 3.13 & 1.17 & 4900 & 1802 & 2.72E+19 & 1.04 & Alkali &\cite{Grunblatt2024} \\ 
         \hline
    \end{tabular}
    \caption{The upper-limit values of $P_{ind}$ and $\delta T$ for all of the $10\,R_\oplus>R_p>4\,R_\oplus$ planets orbiting giant and main-sequence stars with orbital periods $<1000$d from the NASA Exoplanet Archive.  For $P_{ind}$, we calculated the value assuming both the $\eta$ prescription from \protect\cite{Laine2008} and our own alkali metal prescription (described in Section~\ref{ssec:diffusivity}), and quoted the higher $P_{ind}$ value.  We note that the $T_{eq}$ temperatures are calculated assuming full heat redistribution.}
    \label{tab:exo_pind}
\end{table*}

We also performed a similar series of calculations for planets with radii between 2 --- 4\,$R_\oplus$, and were similarly unable to find any planets in this radius range with values of $\delta_T$ above 5\,K.  However, we note that our modeled $\eta$ structure for these exoplanets assumes that they are gas giants, and may not necessarily accurately describe the interiors of smaller bodies with substantial rocky (or icy) cores.

\subsection{Pulsars}

We now consider whether or not inductive heating is a more dominant mechanism around evolved stellar remnants (such as neutron stars), which tend to have larger surface magnetic fields.  While neutron stars can have much higher surface magnetic fields, Figure~\ref{fig:m_power} shows that their total magnetic moments (even in the most extreme cases) tend to be lesser than those of T Tauri stars due to their reduced size.  However, neutron stars can have extremely high rotation rates, which could potentially result in higher values of $P_{ind}$.

The majority of planets we have discovered around stellar remnants orbit pulsars, as pulsar timing gives us a fantastic opportunity to discover and characterize orbiting satellites. Therefore, we can estimate the typical value of $P_{ind}$ on observed pulsar planets and use that to determine whether or not it is substantial compared to the amount of heating from the host.  \cite{Laycock2025} performed a literature review of known pulsar planets, and identified that a significant fraction of claimed pulsar planets may in fact be stripped stellar cores as opposed to being planetary in origin.  However, we will proceed with our calculations assuming all of the listed objects are planets, and simply caution the reader against assuming that these bodies are all traditional planets.

Several of the bodies identified in \cite{Laycock2025} have orbital radii that place them within the Roche lobe of the neutron stars ($p_{orb}<0.4$d).   However, as we only know the minimum masses of pulsar companions, it is possible that many of these bodies are stars, which may have substantially higher densities than Jupiter-sized companions (leading to a lower limiting period).  In addition, several of these bodies have been identified as ``black widow'' systems, in which the host pulsar is actively absorbing mass from its companion (therefore, the systems are not what we would traditionally refer to as ``stable'').

We are interested in estimating the values of $P_{ind}$ for each potential pulsar planet. However, we do not know all of the necessary properties to measure these values.  While \cite{Laycock2025} states the $p_{rot}$, $p_{orb}$, and $M_p \sin(i)$ for each planet, we do not know their radii.  We assigned each planet a radius based on the following criteria:

$$
R_p=\begin{cases}
    0.1, & \text{if } M_p \leq M_\oplus \\
    1, & \text{if } M_\oplus<M_p\leq M_J \\
    3, & \text{if } M_p>M_J
\end{cases}
$$

While these relations are not particularly realistic, we note that we only know minimum masses for these targets, so the actual size of the planet is largely an estimate anyway.  We deliberately chose large radius values in order to estimate upper limits on $P_{ind}$, as we anticipate that $P_{ind}$ is correlated positively with $R_p$ (see Equation~\ref{eqn:phi_star}).

We next assumed that all neutron stars have a radius of $10^4$m and a mass of roughly $1\,M_\odot$.  We estimated the pulsar temperature $T_{\star}$ as being $10^5$\,K, which is comparable to what is observed for typical pulsars \citep{Yanagi2020} \citep[though we note that it is possible for these stars to get cooler- PSR\,J2144-3933, the coldest known neutron star, has a surface temperature of $<42,000$K][]{Guillot2019}. We assign $B_{surf}$ a value of $10^{16}$G, which is comparable to the upper limit possible for neutron stars derived in \cite{Gomes2019}, though we do note that the majority of the \cite{Laycock2025} pulsars with measured surface magnetic fields tend to have much weaker fields than this value \citep[$10^7-10^9$G, see][]{Ng2014, Zhang2025}.  Finally, to get an upper limit on $P_{ind}$, we let $\alpha=\pi/2$ and calculated $P_{ind}$ with both of our $\eta$ prescriptions, keeping the one with the higher value.  We also fixed $e$ at zero.  We had to omit all planets with $T_{eq}<40$K (corresponding roughly to $P>30$d) from our results, as we found that \texttt{GASTLI} could not converge on a interior profile for such cool planets.  Given the strong inverse relationship between orbital period and $P_{ind}$, we consider it unlikely that these long-period planets would have had high values of $P_{ind}$, so omitting these objects is unlikely to result in us missing promising targets.  

With our methods, we identified thirteen systems with upper limit $P_{ind}$ values large enough to cause a $\delta T>1$K- B1257+12, J1719-1438, J2322-2650, J0636+5128, J1311-3430, J1807-2459A, J1653-0158, J0312-0921, J1825-2452G, J1701-3006H, J1928+1245, J0024-7204ac, and J1748-2021H.  For each target, we performed a literature review to improve the accuracy of our parameter estimates so that we could determine whether or not our calculated $\delta T$ values are realistic.  We found that nearly all of these targets either had published magnetic fields \citep{Konacki2003, Ng2014, Zhang2025, Pletsch2012, Lynch2012, Nieder2020, Parent2019, Vleeschower2022} or magnetic fields that could be easily estimated from their spin period and period derivatives using Equation 1.14 from \cite{Caldelano2019} \citep{Smith2023, Begin2006, Vleeschower2024}.  In these cases, the magnetic field was less than $10^{10}$\,G (typically falling between $10^7-10^9$G), far lower than our estimated upper limit.  When updating our calculations to use these more physically reasonable magnetic field values, we found that our new $\delta T <<1$\,K, meaning that we do not anticipate observing meaningful induction heating on these planets, even if they do possess moderate eccentricities.  

The only exception to this trend is PSR\,J0024-7204ac \citep[also known as 47\,Tuc\,ac;][]{Ridolfi2021}, which does not have a measured magnetic field or spin period derivative.  However, we have no reason to believe that this pulsar would be systematically different from any of the other known pulsars with substellar companions.

Therefore, overall, we do not find any evidence that any of the currently observed pulsar planets should show evidence of this heating mechanism.  It appears that $P_{ind}$ is only significant in the most extreme possible conditions- this mechanism can only be relevant if a short-period planet is observed orbiting a magnetar on a polar (or highly eccentric) orbit, as only either magnetar-like magnetic fields or extremely short orbital periods produce extreme values of $P_{ind}$.  

However, as discussed in \cite{Zhang2025}, we currently do not fully understand the mechanisms behind the irradiation of the atmospheres of pulsar planets, which may not necessarily follow the standard $T_{eq}$ calculation.  It therefore may be difficult to definitively pinpoint this mechanism as the cause of heating on these planets, even if we do discover a system where this effect is more likely to occur.

\subsection{White Dwarfs}

Magnetic white dwarfs, as shown in Figure~\ref{fig:m_power}, can have some of the highest values of $m$ for all potential planet hosts, making them superficially appear to be excellent sources of induction heating.  Unfortunately, as they tend to be very dim, there are relatively few white dwarfs with known companions.  The majority of exoplanet-related white dwarf observations involve the presence of disrupted planetary material on the white dwarf's photosphere, which does not enable us to study the effects of magnetic induction.  Upcoming \textit{JWST} observations \citep[such as][]{Vanderburg2025} and analyses of \textit{TESS} products \citep[such as the work discussed in][]{Kunimoto2023} may provide us with improved insight on the occurrence rate of stable white dwarf companions.  Of the white dwarf systems with extant observed companions,  the majority of these systems, such as WD\,0806–661 \citep{Rodrigeuz2011} and KMT-2020-BLG-0414 \citep{Zhang2024} have orbital separations that are far too large for $P_{ind}$ to be significant.  

The best case study we currently have for studying white dwarf planets is WD\,1856+534.  WD\,1856+534 is a white dwarf with a transiting giant planet that has been observed by \textit{JWST} in \cite{Limbach2025}.  The transiting planetary companion has a short enough orbital period that it would have been engulfed by the star during its previous giant phase.  It has therefore been frequently cited as evidence that planets may undergo orbital evolution even after the star enters a post-main sequence phase \citep{Limbach2025, Vanderburg2020}.  In addition, its orbital period is short enough that it may be a candidate for inductive heating.  While the magnetic field and rotation period of WD\,1856+534 are not known, we can estimate an upper limit on $P_{ind}$ by assuming the highest possible value of $B_{surf}$ and the lowest possible value of $p_{rot}$.

\cite{Rosa2024} performed a systematic study of \textit{TESS} data, and measured rotation periods of a large sample of white dwarfs.  The minimum rotation period they measured was that of TIC\,279484490, which had a period of $0.086$d. We will therefore adopt $0.08$d as the value of $p_{rot}$ for our calculations.  We also adopt $B_{surf}$=10$^9$G, in alignment with the fields observed on the highest-field magnetic white dwarfs \citep{Ferrario2015}, and $\alpha = \pi/2$.  We use the \cite{Laine2008} and alkali models for $\eta$, which, as discussed in Section~\ref{sec:comp_to_exo}, may not necessarily accurately describe the conductivity in cooler planets such as WD\,1856+534.  However, we proceed as we have found that $P_{ind}$ is typically not very sensitive to $\eta$.  

We calculated a $P_{ind}$ of $7.9*10^{19}$\,J\,s$^{-1}$, which would result in the planet having a surface temperature of 400\,K- a large increase compared to its zero-albedo $T_{eq}$ of roughly 180\,K. Using the alkali $\eta$ model, we recovered a similar temperature increase, with a surface temperature of 360\,K.  However, the observed effective temperature of WD\,1856+534\,b is $T_{surf}=184\,\pm\,8$\,K \citep{Limbach2025}, meaning that its temperature is not consistent with our calculated value and could, in fact, be explained without any induction heating whatsoever.  

\begin{figure*}
    \centering
    \includegraphics[width=1\linewidth]{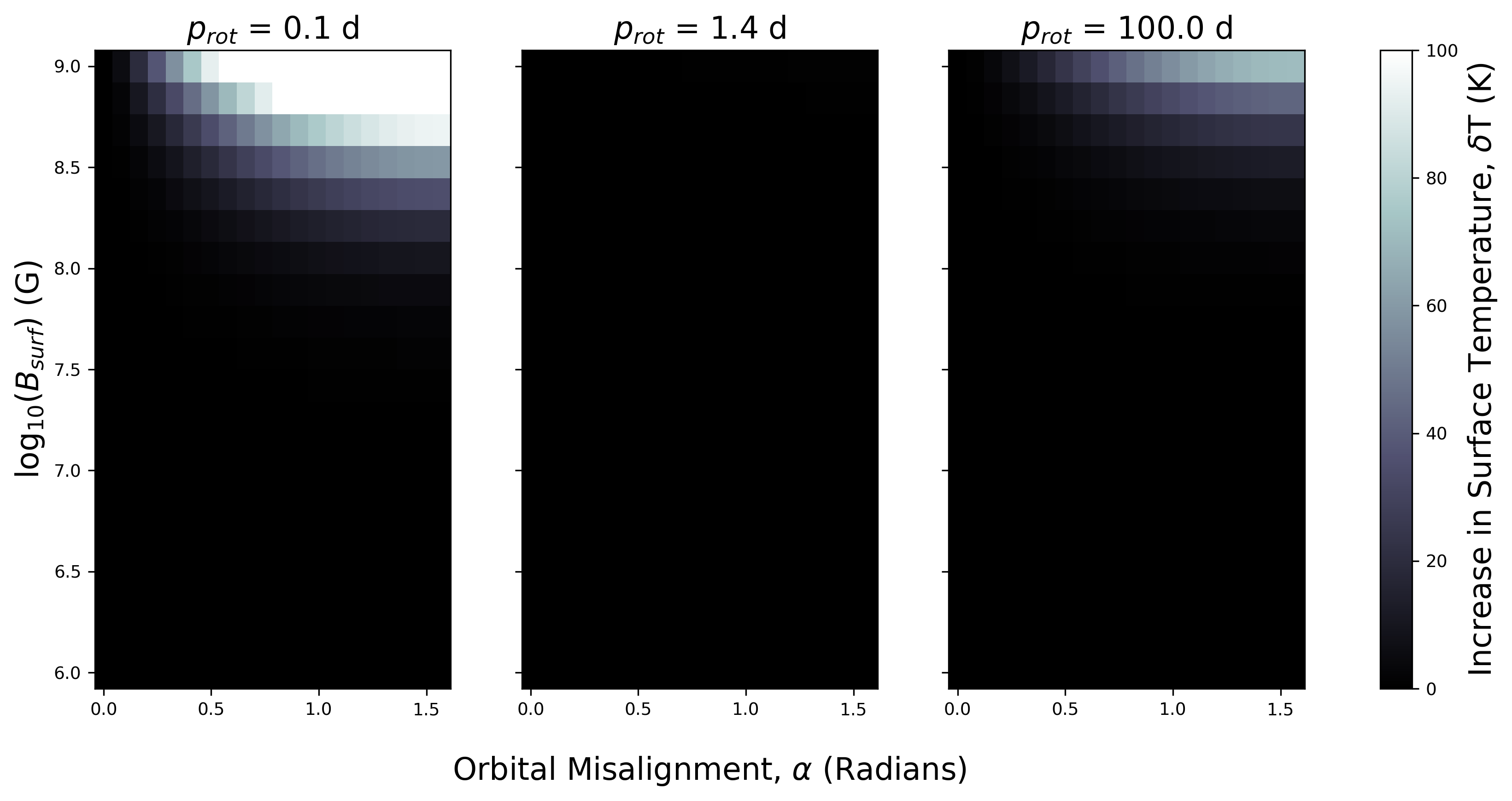}
    \caption{The expected change in surface temperature of WD\,1856+534\,b as a function of its host's and orbital properties.}
    \label{fig:WD_temp_changes}
\end{figure*}

With the measured effective temperature as a prior, we can therefore perform an estimate of the characteristics of the WD\,1856+534 system by determining which combinations of $\alpha$, $b_{surf}$, and $p_{rot}$ result in negligible increases of $T_{surf}$.  Figure~\ref{fig:WD_temp_changes} shows how these different parameters affect the change in surface temperature, $\delta T$, using the \cite{Laine2008} $\eta$ model (which we have found to deliver similar results as the alkali $\eta$ model in this regime).  In general, we see that $\delta T$ is negligible (in this case, $\delta T < 10$) when 1) $\alpha$ is near-zero, 2) $p_{rot} \approx p_{orb}$, and 3) when $B_{surf} < 10^{8}$\,G.  If we are interested in constraining values like $B_{surf}$ by studying the heating of companion planets, it will be necessary to have more precise estimates of the rotation periods of white dwarfs, as well as their obliquities, in the future. 

\section{Conclusions}
\label{sec:conc}

A planet orbiting a star with a misaligned magnetic dipole is expected to experience a time-dependent magnetic field, which in turn could result in the generation of a bulk current on that planet.  Following the work of \cite{Laine2008}, we calculated the power dissipated by this generated field in giant planets in the low Reynolds-number (diffusion-dominated) regime.  We updated their calculations to reflect temperature-pressure profiles from the \texttt{GASTLI} code, as well as including slight updates to fix minor calculation errors. 

We also studied how the magnetically induced power $P_{ind}$ varies with various system parameters.  In general, $P_{ind}$ is an extremely strong function of $m$ ($P_{ind} \propto m^2$), the stellar spin/dipole alignment $\alpha$ ($P_{ind} \propto \sin^2(\alpha)$), and $p_{orb}$ ($P\propto p_{orb}^{-4}$).  However, $P_{ind}$ appears to be far less sensitive to the characteristics of the planet itself, changing by less than an order of magnitude even if the planet's magnetic diffusion $\eta$ or temperature are varied over extreme ranges.  This means that the power induced by the stellar magnetic field is unlikely to create a runaway heating process on these planets. However, the precise value of $\eta$ still holds some importance, as a low value of $\eta$ in the planetary atmosphere could violate our assumption that the Reynolds-number is low.  Our simplified $\eta$ models predict that $\eta$ should be sufficiently high for diffusion to dominate over induction in most scenarios in which the planet has a $>3$d orbital period or in which the planetary wind speed is not superlatively high. However, the effects of unipolar induction are more significant on hot Jupiters with high observed wind speeds.  We leave a more exhaustive study of unipolar induction on giant planets to future works.

We then expanded the calculations of \cite{Laine2008} and \cite{Campbell1983} in order to accommodate the possibility that the orbiting planet has an eccentric orbit.  We found that, to first-order, nonzero eccentricities in planets with higher $P_{ind}$ values, enabling situations in which $P_{ind}>0$ even when the stellar magnetic axis is aligned ($\alpha = 0$).  However, the effect typically does not increase $P_{ind}$ by more than 50\% when $\alpha \neq 0$ and the eccentricity is low.  A more accurate prescription for how higher values of eccentricity influence $P_{ind}$ would require expansions to additional orders of eccentricity, which we leave to future works. We also followed \cite{Chang2012} and studied how changing the planetary obliquity $\lambda$ would influence $P_{ind}$, and found that the effect is similar in magnitude to the relationship between $\alpha$ and $P_{ind}$.

Finally, we estimated upper limits on $P_{ind}$ for the known exoplanet population by calculating $P_{ind}$ for all planets that are likely to have substantial gaseous envelopes, using limits on $m$ informed by a literature search on the magnetic field properties of main sequence stars, giants, white dwarfs, and pulsars.  We then estimated $\delta T$, the increase in surface temperature we would anticipate to see on these planets as a result of the induction heating.  We found that $\delta T$ is far too small to be observable on all of the exoplanets we studied, indicating that $P_{ind}$ is small compared to the power induced by stellar irradiation on all known exoplanets.  

It is, however, theoretically possible for planets with large $\delta T$ values to exist.  These would have to be extremely short-period ($<1d$) planets orbiting superlatively-magnetic stars with misaligned orbits and/or dipole axes.  It is also possible that planets with very high eccentricities may similarly be able to achieve extreme $\delta T$ values, though we cannot accurately calculate the relationship between $P_{ind}$ and eccentricity for high values of eccentricity with our current methods.

\section*{Acknowledgements}

 CEW and DZS acknowledge funding support from JWST GO 5959, which was provided by NASA through a grant from the Space Telescope Science Institute. We thank Stephen Zepf, Adina Feinstein, and Ian Adams for helpful discussions that contributed to this paper, as well as Konstantin Batygin for numerous helpful comments and suggestions.

This research has made use of NASA's Astrophysics Data System Bibliographic Services. 

\textit{Software:} \texttt{GASTLI} \citep{Acuna21, Acuna24}, \texttt{Matplotlib} \citep{matplotlib}, \texttt{Numpy} \citep{numpy}, \texttt{Pandas} \citep{mckinney-proc-scipy-2010, reback2020pandas}, \texttt{Scipy} \citep{2020SciPy-NMeth}, \texttt{Sympy} \citep{Meurer2017}, \texttt{Radvel} \citep{Fulton2017,Fulton2018}

\textit{Facilities:} NASA Exoplanet Archive

%%%%%%%%%%%%%%%%%%%%%%%%%%%%%%%%%%%%%%%%%%%%%%%%%%
\section*{Data Availability}

The data used in this article are included in the NASA Exoplanet Archive, which is accessible \href{https://exoplanetarchive.ipac.caltech.edu/index.html}{online}.  The new code underlying this analysis, including the code used to generate the figures, is available on Github at \texttt{https://github.com/madisontbrady/ohmic\_heating}.

\bibliographystyle{mnras}
\bibliography{example} % if your bibtex file is called example.bib

\appendix

\section{Estimating $C_l^m$}
\label{app:clm}

To calculate $C_l^m$, we can leverage the fact that we anticipate that both $\phi$ and $\partial \phi / \partial r$ will be continuous across the surface of the planet.  In other words, $\phi_{inner} = \phi_p + \phi_\star$ at $r=R_p$.  As $\phi_p$ and $\phi_\star$ are both real, we will need to consider the real components of $\phi_{inner}$.  We can then use the different Legendre polynomials, as well as the trigonometric functions $\cos(\omega t$ and $\sin (\omega t)$ to form a set of bases, which can form the foundations of a system of equations to solve for the $C_l^m$. To more carefully track the real and imaginary components of $\phi_{inner}$, we will explicitly separate the real and imaginary components of $C_l^m$; $C_l^m = \mu_l^m + i \nu_l^m$.

In more detail, we first write out $\phi_{inner}$ in the form:

\begin{equation}
\begin{aligned}
    \phi_{inner} (R_p, t) = \mu_0\sum_{l, m} (\mu_l^m + i \nu_l^m )\bigg(\mathrm{Re}\big(G_l(R_p)\big) + i \mathrm{Im}\big(G_l(R_p)\big)\bigg) \big(\cos(\omega t) + i\sin(\omega t) \big) Y_l^m(\theta, \varphi).
\end{aligned}
\end{equation}

From here forward, we will simplify our writing by omitting the $R_p$ in these equations.  Keep in mind that, from now on, $G_l$ is actually $G_l(R_p)$.

First, we will evaluate the terms of $\phi_{inner}$ for when $l=1$.

\begin{equation}
\begin{aligned}
    \mathrm{Re}\big(\phi_{inner,l=1}\big) = \\ \mu_0 \sqrt{\frac{3}{8\pi}} P_1^1 \Bigg(\cos(\varphi) \cos(\omega t) \bigg(\mathrm{Re}(G_1)\big(\mu_1^{-1} - \mu_1^1 \big) + \mathrm{Im}(G_1)\big(-\nu_1^{-1} +  \nu_1^{1}\big) \bigg) 
    + \sin(\varphi) \cos(\omega t) \bigg( \mathrm{Re}(G_1)\big( \nu_1^{-1} + \nu_1^1 \big) + \mathrm{Im}(G_1)\big(\mu_1^{-1} + \mu_1^1\big) \bigg) \\
    + \cos(\varphi) \sin (\omega t) \bigg(\mathrm{Re}(G_1)\big(-\nu_1^{-1} + \nu_1^1\big) + \mathrm{Im}(G_1)\big(-\mu_1^{-1} +\mu_1^{1}\big) \bigg) 
    + \sin(\varphi) \sin(\omega t) \bigg( \mathrm{Re}(G_1)\big( \mu_1^{-1} + \mu_1^1 \big) + \mathrm{Im}(G_1)\big(-\nu_1^{-1} - \nu_1^1 \big) \bigg) \Bigg)\\ 
    +    \mu_0 \sqrt{\frac{3}{4\pi}} P_1^0 \Bigg( \cos(\omega t) \bigg(\mathrm{Re}(G_1)\mu_1^{0} - \mathrm{Im}(G_1)\nu_1^{0} \bigg) + \sin(\omega t)\bigg(-\mathrm{Re}(G_1)\nu_1^{0}-\mathrm{Im}(G_1)\mu_1^{0}   \bigg)  \Bigg) .
\end{aligned}
\end{equation}

We see here that there are terms that are functions of $P_{1}^0$ and $P_{1}^1$.  However, as we can see in Equations~\ref{eqn:phi_star} and \ref{eqn:phi_p}, there are no field source terms of the form $P_{1}^0$.  Therefore, we anticipate that the $P_1^0$ term in $\phi_{inner}$ also vanishes, and therefore only consider the $P_1^1$ terms.  If we group $P_1^1$ terms from $\phi_{inner}$, $\phi_\star$, and $\phi_p$ in order of the Legendre polynomials and the $\omega t$ and $\phi$ terms, we can create a system of equations:

\begin{equation}
\label{eqn:alpha_bc_first}
\begin{aligned}
   \mathrm{Re}(G_1)\big(\nu_1^{-1}  -\nu_1^1\big) + \mathrm{Im}(G_1)\big(\mu_1^{-1} -\mu_1^{1}\big) + \sqrt{\frac{8\pi}{3}} \frac{1}{R_p}  \alpha_1 =  \sqrt{\frac{8\pi}{3}}\frac{2m \sin(\alpha)R_p^2}{8\pi D^3},
\end{aligned}
\end{equation}

\begin{equation}
\begin{aligned}
    \mathrm{Re}(G_1)\big(-\mu_1^{-1} + \mu_1^1 \big) + \mathrm{Im}(G_1)\big(\nu_1^{-1} -  \nu_1^{1}\big) + \sqrt{\frac{8\pi}{3}}\frac{1}{R_p} \alpha_2  = 0,
\end{aligned}
\end{equation}

\begin{equation}
\begin{aligned}
     \mathrm{Re}(G_1)\big(-\mu_1^{-1} +-\mu_1^1 \big) + \mathrm{Im}(G_1)\big(\nu_1^{-1} + \nu_1^1 \big) + \sqrt{\frac{8\pi}{3}}\frac{1}{R_p} \alpha_3 = 0,
\end{aligned}
\end{equation}

\noindent and:

\begin{equation}
\begin{aligned}
     \mathrm{Re}(G_1)\big( -\nu_1^{-1} + -\nu_1^1 \big) + \mathrm{Im}(G_1)\big(-\mu_1^{-1} + -\mu_1^1\big)   + \sqrt{\frac{8\pi}{3}}\frac{1}{R_p}  \alpha_4 =  \sqrt{\frac{8\pi}{3}}\frac{m \sin(\alpha)R_p^2}{8\pi D^3}.
\end{aligned}
\end{equation}

We can also create four additional constraints assuming the continuity of the derivative at $r=R_p$:

\begin{equation}
\begin{aligned}
   \mathrm{Re}(\dot{G_1})\big(\nu_1^{-1}  -\nu_1^1\big) + \mathrm{Im}(\dot{G_1})\big(\mu_1^{-1} -\mu_1^{1}\big) - \sqrt{\frac{8\pi}{3}} \frac{1}{R_p^2}  \alpha_1 =  \sqrt{\frac{8\pi}{3}}\frac{4m \sin(\alpha)R_p}{8\pi D^3},
\end{aligned}
\end{equation}

\begin{equation}
\begin{aligned}
    \mathrm{Re}(\dot{G_1})\big(-\mu_1^{-1} + \mu_1^1 \big) + \mathrm{Im}(\dot{G_1})\big(\nu_1^{-1} -  \nu_1^{1}\big) - \sqrt{\frac{8\pi}{3}}\frac{1}{R_p^2} \alpha_2  = 0,
\end{aligned}
\end{equation}

\begin{equation}
\begin{aligned}
     \mathrm{Re}(\dot{G_1})\big(-\mu_1^{-1} +-\mu_1^1 \big) + \mathrm{Im}(\dot{G_1})\big(\nu_1^{-1} + \nu_1^1 \big) - \sqrt{\frac{8\pi}{3}}\frac{1}{R_p^2} \alpha_3 = 0,
\end{aligned}
\end{equation}

\noindent and:

\begin{equation}
\label{eqn:alpha_bc_last}
\begin{aligned}
     \mathrm{Re}(\dot{G_1})\big( -\nu_1^{-1} + -\nu_1^1 \big) + \mathrm{Im}(\dot{G_1})\big(-\mu_1^{-1} + -\mu_1^1\big)   - \sqrt{\frac{8\pi}{3}}\frac{1}{R_p^2}  \alpha_4 =  \sqrt{\frac{8\pi}{3}}\frac{2m \sin(\alpha)R_p}{8\pi D^3}.
\end{aligned}
\end{equation}

Therefore, we have 8 constraints and 8 variables ($\mu_1^{-1}$, $\nu_1^{-1}$, $\mu_1^1$, $\nu_1^1$, $\alpha_1$, $\alpha_2$, $\alpha_3$, $\alpha_4$), allowing us to determine $C_{l^m}$ for $\phi_{inner}$ in the $l=1$ case.  We note that we also found what appears to be an error in the equations solving for $\alpha_1, \alpha_2, \alpha_3, \alpha_4$ in Appendix B in \cite{Laine2008}, where the solutions to the derivative equations appear to be mistakenly transposed.  We show their results, alongside our corrections, in Table~\ref{tab:coeff_fixed} We also have some sign differences from the solutions in \cite{Laine2008}, but these are merely due to how we defined the Legendre polynomials and they do not influence our overall results.

\begin{table*}
    \centering
    \begin{tabular}{|cc|}
    \hline
         \cite{Laine2008} & This Paper \\
         \hline

$\mathrm{Re}(\dot{G_1})\big(\nu_1^{-1}  -\nu_1^1\big) + \mathrm{Im}(\dot{G_1})\big(\mu_1^{-1} -\mu_1^{1}\big) + \sqrt{\frac{8\pi}{3}} \frac{1}{R_p^2}  \alpha_1 = 0$ 
&        
$\mathrm{Re}(\dot{G_1})\big(\nu_1^{-1}  -\nu_1^1\big) + \mathrm{Im}(\dot{G_1})\big(\mu_1^{-1} -\mu_1^{1}\big) - \sqrt{\frac{8\pi}{3}} \frac{1}{R_p^2}  \alpha_1 = 2A$
\\
$\mathrm{Re}(\dot{G_1})\big(-\mu_1^{-1} + \mu_1^1 \big) + \mathrm{Im}(\dot{G_1})\big(\nu_1^{-1} -  \nu_1^{1}\big) + \sqrt{\frac{8\pi}{3}}\frac{1}{R_p^2} \alpha_2  = A$
&
$\mathrm{Re}(\dot{G_1})\big(-\mu_1^{-1} + \mu_1^1 \big) + \mathrm{Im}(\dot{G_1})\big(\nu_1^{-1} -  \nu_1^{1}\big) - \sqrt{\frac{8\pi}{3}}\frac{1}{R_p^2} \alpha_2  = 0$
\\
$\mathrm{Re}(\dot{G_1})\big( -\nu_1^{-1} + -\nu_1^1 \big) + \mathrm{Im}(\dot{G_1})\big(-\mu_1^{-1} + -\mu_1^1\big)   + \sqrt{\frac{8\pi}{3}}\frac{1}{R_p^2}  \alpha_4 =  2A$
&
$\mathrm{Re}(\dot{G_1})\big( -\nu_1^{-1} + -\nu_1^1 \big) + \mathrm{Im}(\dot{G_1})\big(-\mu_1^{-1} + -\mu_1^1\big)   - \sqrt{\frac{8\pi}{3}}\frac{1}{R_p^2}  \alpha_4 =  A$
\\  
         \hline
    \end{tabular}
    \caption{The differences between the \protect\cite{Laine2008} equations and the equations used in this paper to solve for $\mu$ and $\nu$ in the $l=1$, $m=1$ scenario. We note that the sign differences in these equations are a result of how we defined the Legendre polynomials and are therefore not physically meaningful.  In this equation, $A = 2 m \mathrm{sin}(\alpha) R_p \sqrt{8\pi/3} / (8\pi D^3)$.}
    \label{tab:coeff_fixed}
\end{table*}

Now, we can consider the (more complicated) $l=2$ case.  Here, we must consider five terms, where $l = 2$ and $m=-2, -1, 0, 1, 2$. We find that:

\begin{equation}
\begin{aligned}
     \mathrm{Re}\big(\phi (R_p, t, l=2)\big) = \\
    \frac{\mu_0}{12}\sqrt{\frac{15}{2 \pi}} P_2^2 \Bigg(
    \cos(2\varphi) \cos(\omega t) \bigg( \mathrm{Re}(G_2) \big( \mu_2^{-2} + \mu_2^2 \big)  + \mathrm{Im}(G_2) \big(-\nu_2^{-2} - \nu_2^2 \big) \bigg)   
    +\sin(2\varphi) \cos(\omega t)\bigg( \mathrm{Re}(G_2) \big(\nu_2^{-2} - \nu_2^2 \big)  + \mathrm{Im}(G_2) \big(\mu_2^{-2} - \mu_2^2 \big) \bigg) \\
    +\cos(2\varphi) \sin (\omega t)\bigg( \mathrm{Re}(G_2) \big(-\nu_2^{-2} - \nu_2^2 \big)  + \mathrm{Im}(G_2) \big(-\mu_2^{-2} - \mu_2^2 \big) \bigg) 
    + \sin (2\varphi) \sin (\omega t) \bigg( \mathrm{Re}(G_2) \big(\mu_2^{-2} - \mu_2^2 \big)  + \mathrm{Im}(G_2) \big(-\nu_2^{-2} + \nu_2^2 \big) \bigg)\Bigg) \\
    + \frac{\mu_0}{2} \sqrt{\frac{5}{\pi}} P_2^0 \Bigg( \cos(\omega t) \bigg(\mathrm{Re}(G_2) \mu_2^0 - \mathrm{Im}(G_2) \nu_2^0  \bigg) 
    + \sin(\omega t) \bigg(-\mathrm{Re}(G_2) \nu_2^0 - \mathrm{Im}(G_2)\mu_2^0   \bigg) \Bigg).
\end{aligned}
\end{equation}

In the $l=2, m=0$ scenario,we can calculate the coefficients...

\begin{equation}
\label{eqn:beta_bc_first}
\begin{aligned}
    -\mathrm{Re}(G_2) \nu_2^0 - \mathrm{Im}(G_2)\mu_2^0    
    - \sqrt{\frac{ 4\pi}{5}} \frac{1}{R_p^2} \beta_1 =  \sqrt{\frac{4\pi}{5}}\frac{ m \sin (\alpha) R_p^3}{8 \pi D^4},
\end{aligned}
\end{equation}

\noindent and:

\begin{equation}
\begin{aligned}
       \mathrm{Re}(G_2) \mu_2^0 - \mathrm{Im}(G_2) \nu_2^0   
   -  \sqrt{\frac{4\pi}{5}} \frac{1}{R_p^2} \beta_2 = 0.
\end{aligned}
\end{equation}

Once again, the continuity of the derivative supplies two additional constraints:

\begin{equation}
\begin{aligned}
    -\mathrm{Re}(\dot{G_2}) \nu_2^0 - \mathrm{Im}(\dot{G_2})\mu_2^0    
    +2 \sqrt{\frac{4\pi}{5}} \frac{1}{R_p^3} \beta_1 =  3 \sqrt{\frac{4\pi}{5}}\frac{ m \sin (\alpha) R_p^2}{8 \pi D^4},
\end{aligned}
\end{equation}

\noindent and:

\begin{equation}
\label{eqn:beta_bc_last}
\begin{aligned}
       \mathrm{Re}(\dot{G_2}) \mu_2^0 - \mathrm{Im}(\dot{G_2}) \nu_2^0   
   +2 \sqrt{\frac{4\pi}{5}} \frac{1}{R_p^3} \beta_2 = 0.
\end{aligned}
\end{equation}

In the $m=\pm2$ scenario, we find:

\begin{equation}
\label{eqn:gamma_bc_first}
\begin{aligned}
    \mathrm{Re}(G_2) \big(-\nu_2^{-2} - \nu_2^2 \big)  + \mathrm{Im}(G_2) \big(-\mu_2^{-2} - \mu_2^2 \big) 
    -12 \sqrt{\frac{2\pi}{15}}\frac{1}{R_p^2} \gamma_1=
        -12 \sqrt{\frac{2\pi}{15}}\frac{ m \sin (\alpha) R_p^3}{8 \pi D^4}  \bigg( \frac{1}{2}  \bigg),
\end{aligned}
\end{equation}

\begin{equation}
\begin{aligned}
     \mathrm{Re}(G_2) \big( \mu_2^{-2} + \mu_2^2 \big)  + \mathrm{Im}(G_2) \big(-\nu_2^{-2} - \nu_2^2 \big) 
    -12 \sqrt{\frac{2\pi}{15}}\frac{1}{R_p^2}  \gamma_2 = 0,  \\
\end{aligned}
\end{equation}

\begin{equation}
\begin{aligned}
    \mathrm{Re}(G_2) \big(\mu_2^{-2} - \mu_2^2 \big)  + \mathrm{Im}(G_2) \big(-\nu_2^{-2} + \nu_2^2 \big)
    -12 \sqrt{\frac{2\pi}{15}}\frac{1}{R_p^2} \gamma_3 = 0 ,
\end{aligned}
\end{equation}
\noindent and:

\begin{equation}
\begin{aligned}
    \mathrm{Re}(G_2) \big(\nu_2^{-2} - \nu_2^2 \big)  + \mathrm{Im}(G_2) \big(\mu_2^{-2} - \mu_2^2 \big) 
  - 12 \sqrt{\frac{2\pi}{15}}\frac{1}{R_p^2}  \gamma_4 =
        -12 \sqrt{\frac{2\pi}{15}}\frac{ m \sin (\alpha) R_p^3}{8 \pi D^4}  \bigg(  \frac{1}{3}  \bigg).\\
\end{aligned}
\end{equation}

The continuity of the derivative here supplies four additional constraints, enabling us to solve for the values of $\mu$ and $\nu$.  

\begin{equation}
\begin{aligned}
    \mathrm{Re}(\dot{G_2}) \big(-\nu_2^{-2} - \nu_2^2 \big)  + \mathrm{Im}(\dot{G_2}) \big(-\mu_2^{-2} - \mu_2^2 \big) 
    +24 \sqrt{\frac{2\pi}{15}}\frac{1}{R_p^3} \gamma_1=
        -12 \sqrt{\frac{2\pi}{15}}\frac{ m \sin (\alpha) R_p^2}{8 \pi D^4}  \bigg( \frac{3}{2}  \bigg),
\end{aligned}
\end{equation}

\begin{equation}
\begin{aligned}
     \mathrm{Re}(\dot{G_2}) \big( \mu_2^{-2} + \mu_2^2 \big)  + \mathrm{Im}(\dot{G_2}) \big(-\nu_2^{-2} - \nu_2^2 \big) 
    +24 \sqrt{\frac{2\pi}{15}}\frac{1}{R_p^3}  \gamma_2 = 0  ,\\
\end{aligned}
\end{equation}

\begin{equation}
\begin{aligned}
    \mathrm{Re}(\dot{G_2}) \big(\mu_2^{-2} - \mu_2^2 \big)  + \mathrm{Im}(\dot{G_2}) \big(-\nu_2^{-2} + \nu_2^2 \big)
    +24 \sqrt{\frac{2\pi}{15}}\frac{1}{R_p^3} \gamma_3 = 0, 
\end{aligned}
\end{equation}
\noindent and:

\begin{equation}
\label{eqn:gamma_bc_last}
\begin{aligned}
    \mathrm{Re}(\dot{G_2}) \big(\nu_2^{-2} - \nu_2^2 \big)  + \mathrm{Im}(\dot{G_2}) \big(\mu_2^{-2} - \mu_2^2 \big) 
  - 12 \sqrt{\frac{2\pi}{15}}\frac{1}{R_p^3}  \gamma_4 =
        -12 \sqrt{\frac{2\pi}{15}}\frac{ m \sin (\alpha) R_p^2}{8 \pi D^4}.  \\
\end{aligned}
\end{equation}

With all of these constraints, we can solve for all of the $C_l^m$ for $l=0,1,2$.  We solved for the unknowns using \texttt{numpy's} \texttt{linalg.solve} and our numerically estimated values of $G_l$.

Overall, our constraints on $C_l^m$ generally agree with those from Appendix B in \cite{Laine2008}, with the major exceptions being some sign conventions (we followed the sign conventions from \cite{Campbell1983} instead) and the errors in the $l=1$ derivatives.

\section{Estimating the $C_l^m$ Coefficients of an Eccentric Planet}
\label{app:clm_ecc}

We can estimate the $C_l^m$ coefficients similarly to how we performed our calculations in the circular-orbit case, but now with additional terms.

First, consider the $l=1, m=0$ term of $\phi_{inner}$:

\begin{equation}
\begin{aligned}
    \phi_{inner, l=1, m=0} (r, t) = \mu_0  C^{0}_{1,1} G_{1,1}(r) Y_1^0(\theta, \varphi)e^{i\omega t}  +  \mu_0 C^{0}_{2,1} G_{2,l}(r) Y_1^0(\theta, \varphi)e^{i\omega_p t} +\\  \mu_0  C^{0}_{3,1} G_{3,1}(r) Y_1^0(\theta, \varphi)e^{i(\omega+\omega_p) t} + \mu_0  C^{0}_{4,1} G_{4,1}(r) Y_1^0(\theta, \varphi)e^{i(\omega-\omega_p) t} .
\end{aligned}
\end{equation}

We are interested in the real portion of this term, as $\phi_\star$ and $\phi_p$ are both real, finding:

%\begin{equation}
%\begin{aligned}
%     \mathrm{Re}(\phi_{inner, l=1, m=0}) = \mu_0 \sqrt{\frac{3}{4\pi}} P_1^0
%     \times \Bigg(\cos(\omega t) \big( \mathrm{Re}(G_{1,1}) \mu_{1,1}^0  -\mathrm{Im}(G_{1,1}) \nu_{1,1}^0)    \big)
%+\sin(\omega t) \big(- \mathrm{Re}(G_{1,1}) \nu_{1,1}^0 - \mathrm{Im}(G_{1,1}) \mu_{1,1}^0   \big)  \\
%+\cos(\omega_p t) \big(  \mathrm{Re}(G_{2,1})\mu_{2,1}^0 - \mathrm{Im}(G_{2,1}) \nu_{2,1}^0 \big)  
%+\sin(\omega_p t) \big(- \mathrm{Re}(G_{2,1})\nu_{2,1}^0 - \mathrm{Im}(G_{2,1}) \mu_{2,1}^0   \big) \\
%+ \cos\big((\omega+\omega_p) t\big) \big(\mathrm{Re}(G_{3,1}) \mu_{3,1}^0 -\mathrm{Im}(G_{3,1}) \nu_{3,1}^0 \big) 
%+ \sin\big((\omega+\omega_p) t\big) \big( -  \mathrm{Re}(G_{3,1}) \nu_{3,1}^0 -  \mathrm{Im}(G_{3,1}) \mu_{3,1}^0 \big) \\
%+ \cos\big((\omega-\omega_p) t\big) \big(\mathrm{Re}(G_{4,1}) \mu_{4,1}^0 -\mathrm{Im}(G_{4,1}) \nu_{3,1}^0 \big) 
%+ \sin\big((\omega-\omega_p) t\big) \big( -  \mathrm{Re}(G_{4,1}) \nu_{4,1}^0 -  \mathrm{Im}(G_{4,1}) \mu_{4,1}^0 \big) \Bigg).
%\end{aligned}
%\end{equation}

\begin{equation}
\begin{aligned}
\label{eqn:ecc_m0l1}
     \mathrm{Re}(\phi_{inner, l=1, m=0}) = \mu_0 \sqrt{\frac{3}{4\pi}} P_1^0 \sum_{i=1}^4
      \Bigg(\cos(\omega_i t) \big( \mathrm{Re}(G_{i,1}) \mu_{i,1}^0  -\mathrm{Im}(G_{i,1}) \nu_{i,1}^0)    \big) +\sin(\omega_i t) \big(- \mathrm{Re}(G_{i,1}) \nu_{i,1}^0 - \mathrm{Im}(G_{i,1}) \mu_{i,1}^0   \big) \Bigg),
\end{aligned}
\end{equation}

\noindent where $\omega_1=\omega$, $\omega_2=\omega_p$, $\omega_3 = \omega+\omega_p$, and $\omega_4 = \omega-\omega_p$.

We can see from our boundary conditions that only the terms that are functions of $\omega_p$ will not vanish (that is, $\mu_{1, 1}^0 = \mu_{3,1}^0= \mu_{4,1}^0= \nu_{1,1}^0= \nu_{3,1}^0= \nu_{4,1}^0 = 0$).  Therefore, we can write out a system of equations to solve for the $i=2$ terms using the boundary conditions:

\begin{equation}
\begin{aligned}
  \mathrm{Re}(G_{2,1})\mu_{2,1}^0 - \mathrm{Im}(G_{2,1}) \nu_{2,1}^0 
-  \sqrt{\frac{4\pi}{3}}\frac{1}{R_p}  \epsilon_1  =  -\sqrt{\frac{4\pi}{3}}\frac{m R_p^2\cos(\alpha)}{8\pi a^3} 3 e ,
\end{aligned}
\end{equation}

\noindent and

\begin{equation}
\begin{aligned}
- \mathrm{Re}(G_{2,1})\nu_{2,1}^0 - \mathrm{Im}(G_{2,1}) \mu_{2,1}^0  
    - \sqrt{\frac{4\pi}{3}}\frac{1}{R_p}  \epsilon_2 = 0.
\end{aligned}
\end{equation}

With the continuity of the derivative, we find two more equations:

\begin{equation}
\begin{aligned}
  \mathrm{Re}(\dot{G_{2,1}})\mu_{2,1}^0 - \mathrm{Im}(\dot{G_{2,1}}) \nu_{2,1}^0 
+  \sqrt{\frac{4\pi}{3}}\frac{1}{R_p^2}  \epsilon_1  =  -2\sqrt{\frac{4\pi}{3}}\frac{m R_p\cos(\alpha)}{8\pi a^3} 3 e ,
\end{aligned}
\end{equation}

\noindent and

\begin{equation}
\begin{aligned}
- \mathrm{Re}(G_{2,1})\nu_{2,1}^0 - \mathrm{Im}(G_{2,1}) \mu_{2,1}^0  
    +\sqrt{\frac{4\pi}{3}}\frac{1}{R_p^2}  \epsilon_2 = 0.
\end{aligned}
\end{equation}

Next, we consider the $m=\pm1, l=1$ case:

\begin{equation}
\begin{aligned}
\mathrm{Re}(\phi_{inner, l=1, m=\pm 1}) = \mu_0 \sqrt{\frac{3}{8\pi}}P_1^1 \sum_{i=1}^4 \Bigg(\cos(\omega_i t) \cos(\varphi)  \bigg(\mathrm{Re}(G_{i,1}) ( -\mu_{i,1}^{1}  + \mu_{i,1}^{-1}) +  \mathrm{Im}(G_{i,1}) (\nu_{i,1}^1 - \nu_{i,1}^{-1})   \bigg)\\
+\cos(\omega_i t) \sin(\varphi) \bigg(\mathrm{Re}(G_{i,1}) ( \nu_{i,1}^1 + \nu_{i,1}^{-1}) +  \mathrm{Im}(G_{i,1}) (\mu_{i,1}^{1} + \mu_{i,1}^{-1})  \bigg) \\
+\sin(\omega_i t)\cos(\varphi) \bigg(\mathrm{Re}(G_{i,1}) (\nu_{i,1}^1-\nu_{i,1}^{-1})  \mathrm{Im}(G_{i,1}) ( \mu_{i,1}^{1}-  \mu_{i,1}^{-1}) \bigg)  \\
+ \sin(\omega_i t) \sin(\varphi) \bigg(\mathrm{Re}(G_{i,1})  (\mu_{i,1}^{1} + \mu_{i,1}^{-1})  + \mathrm{Im}(G_{i,1})  (-\nu_{i,1}^1 - \nu_{i,1}^{-1})  \bigg) \Bigg),
\end{aligned}
\end{equation}

\noindent with $\omega_i$ defined the same way as it was in Equation~\ref{eqn:ecc_m0l1}.  From the boundary conditions, the $i=2$ terms vanish.  The $i=1$ boundary conditions are identical to those in Equations~\ref{eqn:alpha_bc_first}-\ref{eqn:alpha_bc_last}, so we instead reproduce the equations that enable us to solve for the $i=3$ and $i=4$ terms:

\begin{equation}
\begin{aligned}
\mathrm{Re}(G_{4,1}) ( -\mu_{4,1}^{1}  + \mu_{4,1}^{-1}) +  \mathrm{Im}(G_{4,1}) (\nu_{4,1}^1 - \nu_{4,1}^{-1}) 
    - \sqrt{\frac{8\pi}{3}} \frac{1}{R_p}  \xi_1 
    =0,
\end{aligned}
\end{equation}

\begin{equation}
\begin{aligned}
\mathrm{Re}(G_{3,1}) ( -\mu_{3,1}^{1}  + \mu_{3,1}^{-1}) +  \mathrm{Im}(G_{3,1}) (\nu_{3,1}^1 - \nu_{3,1}^{-1})  
    - \sqrt{\frac{8\pi}{3}} \frac{1}{R_p} \xi_2 
    =0,
\end{aligned}
\end{equation}

\begin{equation}
\begin{aligned}
\mathrm{Re}(G_{4,1}) (\nu_{4,1}^1-\nu_{4,1}^{-1}) +   \mathrm{Im}(G_{4,1}) ( \mu_{4,1}^{1}-  \mu_{4,1}^{-1}) 
    - \sqrt{\frac{8\pi}{3}}\frac{1}{R_p}  \xi_3
    =
    -6 e\sqrt{\frac{8\pi}{3}}\frac{ m R_p^2 \sin(\alpha)}{8\pi a^3},
\end{aligned}
\end{equation}

\begin{equation}
\begin{aligned}
\mathrm{Re}(G_{3,1}) (\nu_{3,1}^1-\nu_{3,1}^{-1})  + \mathrm{Im}(G_{3,1}) ( \mu_{3,1}^{1}-  \mu_{3,1}^{-1})
    - \sqrt{\frac{8\pi}{3}}\frac{1}{R_p}  \xi_4 =0,
\end{aligned}
\end{equation}

\begin{equation}
\begin{aligned}
\mathrm{Re}(G_{4,1}) ( \nu_{4,1}^1 + \nu_{4,1}^{-1}) +  \mathrm{Im}(G_{4,1}) (\mu_{4,1}^{1} + \mu_{4,1}^{-1}) 
    - \sqrt{\frac{8\pi}{3}}\frac{1}{R_p}  \xi_5
    =
    -\frac{9}{2} e\sqrt{\frac{8\pi}{3}}\frac{ m R_p^2 \sin(\alpha)}{8\pi a^3}  ,
\end{aligned}
\end{equation}

\begin{equation}
\begin{aligned}
\mathrm{Re}(G_{3,1}) ( \nu_{3,1}^1 + \nu_{3,1}^{-1}) +  \mathrm{Im}(G_{3,1}) (\mu_{3,1}^{1} + \mu_{3,1}^{-1}) 
    - \sqrt{\frac{8\pi}{3}}\frac{1}{R_p}  \xi_6 =
    \frac{3}{2} e\sqrt{\frac{8\pi}{3}}\frac{ m R_p^2 \sin(\alpha)}{8\pi a^3}  ,
\end{aligned}
\end{equation}

\begin{equation}
\begin{aligned}
\mathrm{Re}(G_{4,1})  (\mu_{4,1}^{1} + \mu_{4,1}^{-1})  + \mathrm{Im}(G_{4,1})  (-\nu_{4,1}^1 - \nu_{4,1}^{-1}) 
    - \sqrt{\frac{8\pi}{3}}\frac{1}{R_p}  \xi_7=0,
\end{aligned}
\end{equation}

\noindent and:

\begin{equation}
\begin{aligned}
\mathrm{Re}(G_{3,1})  (\mu_{3,1}^{1} + \mu_{3,1}^{-1})  + \mathrm{Im}(G_{3,1})  (-\nu_{3,1}^1 - \nu_{3,1}^{-1})  \bigg) 
    - \sqrt{\frac{8\pi}{3}}\frac{1}{R_p}  \xi_8 =0.
\end{aligned}
\end{equation}

These conditions, along with the continuity of their derivatives at $r=R_p$, enable us to easily solve for the necessary values of $\mu_{i, 1}^1$, $\mu_{i, 1}^{-1}$, $\nu_{i, 1}^1$, and $\nu_{i, 1}^{-1}$ for $i=3,4$.

\section{Estimating the $C_l^m$ Coefficients of a Planet with Nonzero $\lambda$}
\label{app:clm_lam}

For the $P_1^1$ term, we find:

\begin{equation}
\begin{aligned}
    \mathrm{Re}(G_1)\big(\nu_1^{-1} - \nu_1^1\big) + \mathrm{Im}(G_1)\big(\mu_1^{-1} -\mu_1^{1}\big)  
    +\sqrt{\frac{8\pi}{3}}\frac{1}{R_p} \alpha_1 
    =0,
\end{aligned}
\end{equation}

\begin{equation}
\begin{aligned}
    \mathrm{Re}(G_1)\big(-\mu_1^{-1} + \mu_1^1 \big) + \mathrm{Im}(G_1)\big(\nu_1^{-1} -  \nu_1^{1}\big)
    +\sqrt{\frac{8\pi}{3}}\frac{1}{R_p}  \alpha_2  
    =2\sqrt{\frac{8\pi}{3}} \frac{m  \sin(\lambda) R_p^2}{8 \pi D^3},
\end{aligned}
\end{equation}

\begin{equation}
\begin{aligned}
   \mathrm{Re}(G_1)\big( -\mu_1^{-1} - \mu_1^1 \big) + \mathrm{Im}(G_1)\big(\nu_1^{-1} + \nu_1^1 \big) 
    +\sqrt{\frac{8\pi}{3}}\frac{1}{R_p} \alpha_3
    =\sqrt{\frac{8\pi}{3}} \frac{m  \sin(\lambda) R_p^2}{8 \pi D^3} ,
\end{aligned}
\end{equation}

\noindent and:

\begin{equation}
\begin{aligned}
  \mathrm{Re}(G_1)\big( -\nu_1^{-1} - \nu_1^1 \big) + \mathrm{Im}(G_1)\big(-\mu_1^{-1} - \mu_1^1\big)
    +\sqrt{\frac{8\pi}{3}}\frac{1}{R_p}\alpha_4 
    =0.
\end{aligned}
\end{equation}

These conditions, and their derivatives, can be used to solve for $C_l^m$.  

\bsp	% typesetting comment
\label{lastpage}
\end{document}